\documentclass[10pt,letterpaper]{article}

\usepackage{amssymb}
\usepackage{latexsym}
\usepackage[dvips]{graphics,lscape,rotating}
\usepackage{euscript}
\usepackage{bbm}
\usepackage{amsfonts,amsmath,amssymb}
\usepackage{mathrsfs}
\usepackage{array,longtable}
\usepackage{pstricks,psfrag}
\usepackage{cancel}

\usepackage{makeidx}
\makeindex

\numberwithin{equation}{section}

\newcommand{\mb}[1]{\mathbbm{#1}}

\newcommand{\ol}[1]{\overline{#1}}

\newcommand{\NN}{\nonumber\\}

\newcommand{\barr}[1]{\begin{eqnarray}\begin{array}{#1}}
\newcommand{\earr}{\end{array}\end{eqnarray}}

\newcommand{\WT}[1]{\widetilde{#1}}

\newcommand{\MC}[1]{\mathcal{#1}}
\newcommand{\cc}[1]{\overline{#1}}

\newcommand{\nablaA}{\nabla^A}

\def\e{\varepsilon}
\def\i{\mathbbm{i}}

\def\M{\MC{M}}

\def\MT{\widetilde{M}}
\def\K{\mathcal{K}}
\def\l{\ell}

\def\Xl{X^{(\ell)}}
\def\dXl{\dot{X}^{(\ell)}}
\def\ddXl{\ddot{X}^{(\ell)}}

\def\ba{\begin{eqnarray}}
\def\ea{\end{eqnarray}}
\def\be{\begin{equation}}
\def\ee{\end{equation}}

\newtheorem{Theorem}{Theorem}[section]

\newcommand{\Ho}{\mathcal{H}_{0}}

\DeclareMathOperator{\tr}{tr}

\newtheorem{definition}{Definition}
\newtheorem{lemma}{Lemma}

\begin{document}

\title{\bf{Symmetry Reduction of Loop Quantum Gravity}}
\author{Johannes Brunnemann\\
\texttt{johannes.brunnemann@math.upb.de}\\
Department of Mathematics, University of Paderborn\\
33098 Paderborn\\ Germany\\
{\it{and}}\\
Tim A. Koslowski\\ \texttt{tkoslowski@perimeterinstitute.ca}\\
Perimeter Institute for Theoretical Physics,\\
31 Caroline Street North,\\ Waterloo, Ontario N2J 2Y5\\Canada}
\date{\today}
\maketitle

\begin{abstract}
  The relation between standard Loop Quantum Cosmology and full Loop Quantum Gravity fails already at the first nontrivial step: The configuration space of Loop Quantum Cosmology can not be embedded into the configuration space of full Loop Quantum Gravity due to a topological obstruction. We investigate this obstruction in detail, because many topological obstructions are the source of physical effects. For this we derive the topology of a large class of subspaces of the Loop Quantum Gravity configuration space. This allows us to find the extension of the standard Loop Quantum Cosmology configuration space that admits an embedding in agreement with \cite{Fleischhack2010}. We then construct the embedding for flat FRW Loop Quantum Cosmology and find that it coincides asymptotically with standard LQC.
\end{abstract}
\tableofcontents
\section{Introduction}
The construction of a UV-complete theory of quantum gravity is one of the main open problems in theoretical physics, that sparked research programs in a variety of directions. A particularly fruitful program is Loop Quantum Gravity (LQG) \cite{Thiemann2007,Ashtekar:2004eh,Rovelli:2004tv}, as it represents a mathematically well developed quantum field theoretic framework. An important source of insight for the development of LQG is Loop Quantum Cosmology (LQC) which is a symmetry reduced quantum mechanical model that exhibits many features of full LQG. The expectation that LQC captures essential features of LQG is often motivated by the following argument: On the one hand one expects that quantum effects are most important near a singularity of classical General Relativity (GR); on the other hand one expects in light of the BKL conjecture \cite{Belinsky:1970ew,Khalatnikov:1969eg,Belinsky:1982pk} that the dynamics near a singularity of GR is well approximated by the dynamics of decoupled homogeneous cosmologies, so one may expect that LQC provides essential insight for the dynamics of full LQG in situations where the quantum effects are expected to become most significant.
 
Symmetry reduced models of LQG, in particular LQC, are obtained as \lq\lq loop quantizations\rq\rq of symmetry reductions of the classical theory underlying LQG. This \lq\lq classical reduction then quantization\rq\rq procedure however weakens the link between the reduced and full theory, because important features of the full quantum field theory may be overlooked by going through a classical symmetry reduction and re-quantization. Ideally, one

A phenomenological investigation of this issue has been avoided in the LQC framework and the Spin Foam Cosmology (SFC) framework (see e.g. \cite{Rovelli:2008,Rovelli:2010}). It can be easily understood why SFC are insensitive: standard Spin Foams are not a path integral version of LQG. They are from the point of view of canonical LQG a restriction of full LQG to piece-wise linear LQG with low-valent vertices and unknotted edges. However, standard LQC is embeddable into full piecewise linear LQG\footnote{\label{footnote:plLQG} Piecewise linear LQC \cite{Engle2010} is an interesting modification of standard Loop Quantum Gravity, where arbitrary graphs are replaced by piecewise linear graphs, where edges can be knotted and vertices can possess arbitrarily high valence. The kinematic theory differs significantly form standard Loop Quantum Gravity, while the diffeomorphism-invariant theory can be shown to be equivalent to diffeomorphism-invariant standard LQG if a particular extension of the diffeomorphism group is chosen. However, the precise construction of the path groupoid and extension of the diffeomorphism group may have observable consequences and it is thus {\bf not} for a theorist {\bf but} for experiment to decide which construction if any is realized in nature.}. The embeddability issue is thus avoided at the very start of the program. This is the technical reason why the investigation in this paper is independent of the significant advances in SPC. 

Canonical LQC is on the other hand sensitive to the issue. It seems that the standard LQC avoided the issue so far because only the topological obstruction to embeddability is known and the work-around in terms of piecewise linear LQG exists (we will comment on this at the beginning of section 5.2); however exploring possible deviations from standard LQC should be worthwhile. It is the purpose of this paper perform the first step into this direction. For this it is first necessary to investigate the topological origin of the obstruction to embedding standard LQC into LQG in detail and then to provide an explicit construction of an extension of standard LQC that permits an embedding. The investigation of this mathematical problem is physically motivated by the fact that configuration space topology is the source of interesting effects e.g. in solid state physics or in Euclidean QFT. Moreover, one of the most celebrated results of full LQG is the kinematic discreteness of the area operator. This can be understood as a direct consequence of the topology of the configuration space. We can in light of these precedences not a priori exclude that an embeddable version of LQC admits a very different phenomenology than standard LQC. The investigation of this phenomenology is however a vast subject and is thus beyond the scope of this paper. The main results of this paper are
\begin{enumerate}
 \item The induced topology on one-dimensional affine subspaces of the configuration space of full LQG is the spectrum of the algebra of continuous asymptotically almost periodic functions. This topology is finer than the Bohr compactification of the group $(\mathbb R,+)$, which is used to model the compactness of the configuration space of full LQG in standard LQC.
 \item We construct an embeddable version of flat FRW LQC, which differs from standard LQC by the introduction of configuration operators that vanish at infinity. However, standard LQC and embeddable LQC coincide asymptotically. Standard LQC thus captures the universal holonomy modifications.
 \item We do not see any obstruction to generalizing our construction of an embeddable flat FRW LQC to other models such as \cite{Ashtekar:2009}.
\end{enumerate}
The paper is structured as follows: We first perform necessary asymptotic analysis in sections 2-4, which can be skipped by a physically interested reader. The content of the individual sections is as follows:

In section \ref{Setup} we reexamine the issue of non-embeddability of the standard LQC configurations space into the configuration space of full LQG. As found in \cite{Brunnemann:2007du}, the obstruction to embedability of standard (Bianchi I, isotropic) LQC is the violation of almost periodic dependence of general LQG spinnetwork functions when evaluated on the minisuperspace variables used in LQC. As it turns out \cite{Fleischhack2010} the solution is to extend the standard configuration space of LQC \cite{Bojowald2008} to cylindrical functions, which are supported on arbitrary edges, not only on edges which are straight with respect to the metric of a chosen cosmological background. From the mathematical side the problem of constructing this extension is equivalent to solving the differential equation for the parallel transport (in this context also referred to as holonomy) of the Ashtekar connection along arbitrary edges. As usual on minisuperspace the Ashtekar connection can be parametrized by a real parameter $c$. As a warm up we apply the first obvious approach. That is, we apply slight perturbations to straight edges. This leads to a series expansion of the general solution into powers of a perturbation parameter $\epsilon$, where by construction the zeroth order is standard LQC construction. Demanding the perturbation contributions to remain small for arbitrary large values of $c$ turns out to require $\epsilon\sim c^{-1}$. Hence the perturbation approach already hints to a solution where contributions of perturbations lead to corrections as inverse powers of $c$.         
 
In section \ref{LG} we use the Liouville-Green-Ansatz \cite{Olver} to determine the the dependence of an arbitrary spin network function on the minisuperspace variables. For this the general solution to the holonomy ODE in terms of a series expansion in inverse powers of $c$ is constructed to arbitrary finite order in $c^{-1}$. We provide an explicit finite upper bound for the error of truncating the series at arbitrary finite order.

In section \ref{Asymptotics} the limit $c\rightarrow\infty$ of the constructed solution is analyzed. It is shown to coincide with the standard LQC construction. We show that the desired extension of the LQC configuration space consists of the standard part of functions almost periodic in the parameter $c$ plus functions which vanish for $c=0$ and $c\rightarrow\infty$ as observed in \cite{Fleischhack2010}.  

In section \ref{SymmQuantCon} we use this result, to construct an explicit embedding of isotropic Bianchi I LQC into LQG and discuss its relation with standard LQC. 

In section \ref{Conclusions} we conclude with remarks on the physical interpretation of this result. We close our presentation in section \ref{Outlook} with an outlook on future work.

To complete our presentation, the appendix gives parts of the explicit computations. Additionally an alternative derivation for the geometric interpretation of  parameter $c$ in terms of scalar curvature is presented, which uses a recently developed coordinate-free description \cite{Levermann2009} for the Ashtekar variables.

\section{\label{Setup}Setup}
In this section we give a brief introduction to the symmetric setup used for the construction of standard LQC \cite{Bojowald2008}. In appendix \ref{Geometry Ashtekar Vars} a more detailed coordinate free treatment due to \cite{Levermann2009} is provided.

\subsection{\label{Holonomy ODE for Homogeneous Isotropic Cosmological Model}Holonomy ODE for Homogeneous Isotropic Cosmological Model}

Assume a $3+1$-foliation of a globally hyperbolic four dimensional space time $\M\sim \mb{R}\times\Sigma$, where $\Sigma$ denotes 3-dimensional spatial Cauchy surfaces. Let an edge $e\subset \Sigma$ be given. Assume a trivial principal $SU(2)$ fibre bundle and choose a coordinate chart covering a subset of $\Sigma$ containing $e$. Then then pull-back of the real Ashtekar-Barbero connection as a $su(2)$-valued one form to $\Sigma$ can locally be written as $A(x)=A^I_a(x)~dx^a\otimes \tau_I$, where $\{\tau_I\}_{I=1,2,3}$ denotes a basis of $su(2)$. The embedded edge $e$ can be written as a map $e:\mb{R}\supset [0,T]\ni t\mapsto e(t)\subset\Sigma$. The parallel transport of $A$ along the edge $e$ is referred to as its holonomy and is defined by the following ODE (we follow the conventions of \cite{Brunnemann:2007du}). 
\be\label{Setup 1}
  \frac{d}{dt}h\big(e(t)\big)=-A\big(e(t)\big)\,h\big(e(t)\big)
  ~~~\text{with initial condition  }
  h\big(e(0)\big)=\mb{1}_{SU(2)}
\ee
and $A\big(e(t)\big)= A^I_a\big(e(t)\big)~\dot{e}^a(t)\otimes \tau_I $. In what follows we will denote derivatives with respect to $t$ by dots (e.g. $\dot e(t):=\big(\dot{e}^a\partial_a\big)(t)=\frac{d}{dt} e(t)$. Also we frequently suppress the dependence on $t$ to shorten our notation. Hence we write for (\ref{Setup 1})
\[
   \dot{h}=-A(e) h~.
\]
In this paper, we will be particularly concerned with homogeneous cosmology. This is obtained through a simple transitive action of a 3-dimensional Lie-group $G$ on $\Sigma$, which allows the identification of $\Sigma$ with $G$. Using this identification an invariant basis $\{{\bf x}_a\}_{a=1,2,3}$ in $T\Sigma$ can be described by left / right-invariant vector fields on $G$. If one follows an integral curve $\mathcal{I}$ of one of these vector fields on $G$ respectively $\Sigma$, then the components of the metric tensor are constant along that curve:
\[
   \big<{\bf x}_a, {\bf x}_b\big>\big|_\mathcal{I}=g_{ab}\big|_\mathcal{I}=const
\]
Given the invariant basis $\{{\bf x}_a\}_{a=1,2,3}$, the dual invariant basis of $T^*\Sigma$,  $\{{\bf x}^b\}_{b=1,2,3}$, can be obtained from its definition ${\bf x}^b({\bf x}_a)=\delta^b_a$. For Bianchi I, we use $G=\mathbb R^3$. We can furthermore impose isotropy by enlarging the isometry group to $E(3)$. Using a general result on symmetric connections \cite{Brodbeck:1996ma}, one obtains a parametrization of isotropic Bianchi I connections by
$ A=c\cdot \delta^I_a\,dx^a\otimes \tau_I$. Using $\dot e= \dot e^a(t)~\partial_a$ we find $  A(e)= c\cdot \delta^I_a~\dot{e}^a\otimes  \tau_I$. Using the defining representation of $SU(2)$, the holonomy $h$ can be written in matrix form
\[\begin{array}{lclll}
    &~~~~~~~~&h=\left(\begin{array}{rcr}
                               a&~&b\\
                               -\cc{b} && \cc{a}
                             \end{array}\right)
                  ~~~\text{with   }|a|^2+|b|^2=1
\end{array}\]
Taking the usual basis of $su(2)$
\[\begin{array}{lclll}
     \tau_1= \mb{i} \left(\begin{array}{rr}
                                            0 &-1\\
                                     -1 & 0
                                    \end{array}\right)~~~~
                     \tau_2= \mb{i} \left(\begin{array}{rr}
                                            0 &\mb{i}\\
                                           -\mb{i} &~~ 0
                                    \end{array}\right)~~~~
                     \tau_3= \mb{i} \left(\begin{array}{rr}
                                            -1 &~~0\\
                                     0 & 1
                                    \end{array}\right)~~~~ ,
\end{array}\]
and using the shorthands
\[\begin{array}{lclll}
   \dot{e}^1(t)=:\dot x
    \\
                   \dot{e}^2(t)=:\dot y   ~~~~~~~~m=\dot x -\mb{i} \dot y
    \\             \dot{e}^3(t)=:\dot z   ~~~~~~~~n=\dot z
\end{array}\]
we can write (\ref{Setup 1}) as:
\ba
  \left(\begin{array}{rr}
     \dot a &~~\dot b\\
     -\dot{\cc b}  & \dot{\cc a}
      \end{array}\right)
   &=&-\mb{i}\,c
    \left(\begin{array}{rr}
      n &~~ m\\
     {\cc m}  & -n
      \end{array}\right)
    \left(\begin{array}{rr}
      a &~~ b\\
     -{\cc b}  & {\cc a}
      \end{array}\right)   
\ea
with initial conditions: $a(0)=1~,~b(0)=0~.$
From that we obtain two first order ODE's
\ba\label{first order holonomy ODE system}
  \dot a &=& \mb{i}\,c~(na - m\cc b)  \NN
  \dot b &=& \mb{i}\,c~(nb + m\cc a)  ~~~.
\ea
This can be transformed into a second order ODE for $a$.
\barr{lcllllll}\label{ODE 2nd order a}
\ddot{a}&=&N~ a
           + M~\dot{a}~&~~~~~~&\text{with~~}&M&  :=&  \frac{\dot m}{m}\\
                      &&&&\text{and~~}&N&:=&\mb{i}c\big(\dot{n}-Mn+\mb{i}c(m\cc m +n^2)\big)
\earr
where the initial condition for $a$ now read as:
\be\label{ODE 2nd order a initial conditions}
   a(0)=1~~~~\dot a (0)=\mb{i}c~n(0)
\ee
Using the transformation
\be\label{ODE 2nd order subst}
   d=\frac{a}{\sqrt{m}}~~~~~~~~~~~~~~~\fbox{$m\ne 0$}
\ee
we rewrite (\ref{ODE 2nd order a}) to
\be\label{ODE 2nd order d}
   \ddot{d}=\Big\{\frac{1}{4} M^2-\frac{1}{2}\dot{M} +N\Big\}~d
\ee
Equation (\ref{ODE 2nd order d}) is a linear ODE of second order. Its solution $d(t)$ consists of a linear combination of two fundamental solutions $d^{(+)}(t)$ and $d^{(-)}(t)$, that is
\[
  d=d^{(+)}+d^{(-)}  ~~.
\]
With this, the solution to (\ref{ODE 2nd order a}) can be written as
\be\label{Ansatz general solution for a}
   a=\sqrt{m}~\big(A_{(+)}\, d^{(+)}+A_{(-)}\,d^{(-)}\big)~~,
\ee
where $A_{(+)}, A_{(-)}$ are constants to be chosen such that the initial conditions (\ref{ODE 2nd order a initial conditions}) are satisfied. 

\subsection{\label{Non-Embedability of Configuration Spaces}Non-Embedability of Configuration Spaces}

In \cite{Brunnemann:2007du} it was shown, that the configurations space of LQC, given by $\ol{\mb{R}}_{\text{Bohr}}$, the Bohr compactification of the real line cannot be continuously embedded into the configuration space of LQG, the space $\ol{\MC{A}}$ of generalized connections. It was found that in order to retain continuity it is necessary that the solutions (\ref{Ansatz general solution for a}) depend almost periodic on the parameter $c$ for {\it arbitrary} edges, not only edges which are straight with respect to the symmetric background. However it was shown explicitly that generically this is not the case.   

\paragraph{Spiral Arcs.}

In \cite{Brunnemann:2007du} an exact solution to the ODE (\ref{ODE 2nd order a initial conditions}) it was found for spiral edges. This is the most general case in which the ODE has constant coefficients, because in the spiral case $M(t)=M(0)=:M_0$ and $n(t)=n(0)=:n_0$ are constants. Anticipating the notation of section \ref{LG} the ODE (\ref{LG-Vorbereitung}) reads
\be
   \ddot{d}=\kappa^2 \Lambda^2 d~~~~~~~\text{with}~~~\Lambda^2=\Lambda^2(\kappa):=1+\frac{\alpha_0}{\kappa}+\frac{\beta_0}{\kappa^2},
\ee 
where $\kappa:=\mb{i}\,c$, $\alpha_0:=-M_o\,n_0$ and $\beta_0:=\frac{1}{4}M_0^2$ and we have used arc-length parametrization $|m|^2+n^2=1$. It has two fundamental solutions 
$   d^{(\l)}(\kappa,t)=\mb{e}^{\l\kappa\Lambda t }$ and we can obtain a solution to the holonomy ODE (\ref{ODE 2nd order a initial conditions}) and to the original ODE system (\ref{first order holonomy ODE system}) by analogy with section \ref{Power-Series Solution to Holonomy ODE} by imposing initial conditions ($a(\kappa,0)=1,\dot a(\kappa,0)=\kappa n_0$, $b(\kappa,0)=0,\dot b(\kappa,0)=\kappa m(0)$). The solution\footnote{We denote it by German type letters.} is given by
\ba \label{final ODE solution spiral Case 1}
   \mathfrak{a}(\kappa,t)
   &=&\frac{1}{2}\sqrt{\frac{m(t)}{m(0)}}\sum_{\l=\pm 1}
   \left\{\bigg[1 +\l\cdot \frac{2\,\kappa\, n_0-M_0}{2\,\kappa\, \Lambda} \bigg]
   \mb{e}^{\l\kappa t \Lambda}\right\}
   \NN
   \mathfrak{b}(\kappa,t)
   &=&\sqrt{m(t)\,m(0)}\sum_{\l=\pm 1}
   \left\{\l\cdot\frac{1}{2\Lambda}~
   \mb{e}^{\l\kappa t\, \Lambda}\right\}~~~.
\ea
To obtain an explicit parametrization for the solutions (\ref{final ODE solution spiral Case 1}), we set $m(t)=\mu\cdot\mb{e}^{\mb{i}\lambda t}$ (w.l.o.g. $\mu\in\mb{C},\lambda\in\mb{R}$), hence $M_0=\mb{i}\lambda$ and $m(0)=\mu$. Moreover let $n_0:=\nu$ with $\nu\in\mb{R}$, $ |\mu|^2+\nu^2=1$ and $\Delta:=\sqrt{c^2\Lambda^2}=\big[\frac{\lambda^2}{4}+c(c-\nu\lambda)\big]^{1/2}$.   

\begin{figure}[htbp]
\center
\includegraphics{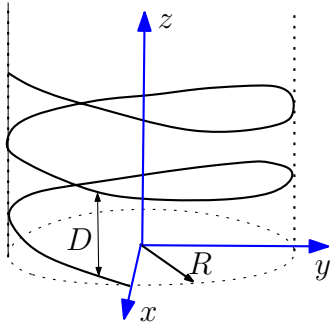}
\caption{Spiral arc as described by our setup. Recall that $m=\dot x -\mb{i} \dot y$ and $n=\dot z$. We have $R^2=x^2+y^2=\frac{|\mu|^2}{\lambda^2}$ and the step height per period is given by $D=\frac{2\pi\nu}{\lambda}$. The limit $\lambda\rightarrow 0$ corresponds to a line in the $(x,y)$-plane with direction $(\Re(\mu),-\Im(\mu))$. The limit $\lambda\rightarrow \infty$ corresponds to a line in $z$-direction. In case of a planar circle ($\nu=0$) we have $R^2=\lambda^{-2}$.    }
\label{Spiral}
\end{figure}

Then (\ref{final ODE solution spiral Case 1}) reads
\ba \label{final ODE solution spiral Case 2}
   \mathfrak{a}(\kappa,t)
   &=&\mb{e}^{\frac{\mb{i}}{2}\lambda t}
   \left\{\cos(\Delta t)+\frac{\mb{i}}{2\Delta}(2\nu c -\lambda)\sin(\Delta t)\right\}
   \NN
   \mathfrak{b}(\kappa,t)
   &=& \mu~\mb{e}^{\frac{\mb{i}}{2}\lambda t}~
   \frac{\mb{i}c}{\Delta}~\sin(\Delta t)~~~.
\ea
Obviously (\ref{final ODE solution spiral Case 2}) exhibits a non-almost dependence of $\mathfrak{a}(\kappa,t), \mathfrak{b}(\kappa,t)$ on $c$, unless the underlying curve is a line (in this case $\lambda=0$ and $\nu=0$). However one observes from (\ref {final ODE solution spiral Case 2}) that in the limit $c\rightarrow \infty$ periodicity is {\it asymptotically} restored. This observation is the starting point for looking at the properties of the holonomies along {\it arbitrary} edges. The first obvious Ansatz is to look at the effect of  small perturbations of straight edges.

\paragraph{Perturbation of Straight Edges}
The detailed computation can be found in section \ref{Perturbation Approach}.  Here we only quote the result. The general setup is given as follows: Let 
$e_0(t)=\big({e_0}^1(t),{e_0}^2(t),{e_0}^3(t)\big)=:\big(x_0,y_0,z_0\big)$. Now assume the edge $\gamma_0$ is deformed into another edge $\gamma$  such that
$\gamma(t)=\big(x_0+\varepsilon\, \WT{x},~y_0+\varepsilon\, \WT{y},~z_0+\varepsilon\, \WT{z}\big)$, where $ \WT{x}= \WT{x}(t),~\WT{y}= \WT{y}(t),~\WT{z}= \WT{z}(t),  $ and  $\varepsilon=const$ is a small deformation parameter. For $e_0$ being a straight line and under some simplifying assumptions described in section \ref{Perturbation about a Line} one obtains a solution to (\ref{Ansatz general solution for a}) in terms of a formal power series in the perturbation parameter $\varepsilon$. However one finds from the computation in section \ref{Perturbation Approach} that in order to ensure \lq\lq small\rq\rq contributions from the perturbation even for large $c$ one has to require that $\varepsilon\sim c^{-1}$. One then obtains a solution to (\ref{Ansatz general solution for a}) which looks as follows (see section \ref{Perturbation about a Line} for computational details):
\be\label{perturbation ansatz result}
     a(\kappa,t)=\sum_{\sigma=\pm}
     \mb{e}^{\sigma\,\kappa\, t} \Big\{
   1+ \mathcal{O}(c^{-1})  \Big\}~~.
\ee
From (\ref{perturbation ansatz result}) it becomes obvious that the property of asymptotic almost periodicity carried by the solutions (\ref {final ODE solution spiral Case 1}) is not a special property of considering spiral arcs. Rather it is a {\it generic} property valid for {\it general}  perturbations of straight edges. Hence we are referred to look for a general solution of (\ref {final ODE solution spiral Case 1}) for large $c$ in terms of an inverse power series of $c$. Fortunately, there is well developed mathematical framework available (see e.g. \cite{Olver}), which uses the so called \lq\lq Liouville Green Ansatz\rq\rq. This is employed in the next section.   

\section{\label{LG}Solution by a Liouville Green Ansatz}
\subsection{Liouville Green Ansatz}
In this section we are going to compute an expression for the holonomy along arbitrary edges. It is based on the presentation  in \cite{Olver}. In principle, this approach is viable for  general differential equations of the form 
\be
   \ddot{a}(\kappa,t)=\kappa^{2n}f(\kappa,t)~a(\kappa,t)
\ee
where $n\in\mb{N}$, $\kappa\in\mb{C}$ is a complex parameter and $f(\kappa,t)$ is an analytic function of $t\in\mb{C}$ with uniform asymptotic expansion 
\be
   f(\kappa,t)=f_0(t)+\frac{f_1(t)}{\kappa}+\frac{f_2(t)}{\kappa^2}+\ldots
   ~~~~~\text{for~~}|\kappa|\rightarrow\infty.
\ee
Using (\ref{ODE 2nd order a}) we can rewrite equation (\ref{ODE 2nd order d}) as
\barr{lcrcl}\label{LG-Vorbereitung}
   \displaystyle\ddot{d}(\kappa,t)=\kappa^2\left(1+\frac{\alpha(t)}{\kappa}+\frac{\beta(t)}{\kappa^2}\right)d(\kappa,t)
   &~~~~~\text{with}~&1&=&f_0(t)=\ol{m}(t)\,m(t)+n(t)^2
   \\
   &&\alpha(t)&:=&f_1(t)=\dot{n}(t)-M(t)\,n(t)   
   \\[1mm]
   &&\displaystyle \beta(t)&:=&f_2(t)=\frac{1}{4}M^2(t)-\frac{1}{2}\dot{M}(t)
   \\[3mm]
   &&&&f_k(t)=0~~~\text{if~}\fbox{$k>2$}~~,
\earr
and $t\in\mb{R}$, $\kappa=\mb{i}c$ and we have chosen arc-length parametrization in order to set $\ol{m}(t)\,m(t)+n(t)^2=1~$. For shortness of the notation we will from now on suppress the dependence on $\kappa,t$ where possible.
\\~\\
Now the strategy is to write (\ref{LG-Vorbereitung}) as 
\be\label{LG-Vorbereitung 2}
   \ddot{d} -\kappa^2\,d= \left(\kappa \alpha+\beta\right)d
\ee
and to regard the rhs as an inhomogeneity. Then two fundamental solutions $d^{(\pm)}(\kappa,t)$ to (\ref{LG-Vorbereitung 2}) can be obtained from the Liouville-Green-Ansatz
\be\label{LG-Ansatz}
   d^{(\l)}(\kappa,t)=\mb{e}^{\l\,\kappa t}X^{(\l)}(t)\left(\sum_{r=0}^\infty\frac{A_r^{(\l)}(t)}{\kappa^r}\right)~~~,
\ee
where $\l=\pm 1$ and $A^{(\l)}_0(t)=const=1$.
Plugging this Ansatz into (\ref{LG-Vorbereitung 2}) we obtain
\barr{rclll}\label{equating coefficients of kappa}
   \ddot{d}^{(\l)}&=&\mb{e}^{\l\,\kappa t}\Big\{&
   \kappa^2 \Xl +\kappa \big(\Xl A_1^{(\l)}+ 2\l\dXl\big)
                +       \big(\Xl A_2^{(\l)}+2\l(\dXl A_1^{(\l)}+\Xl \dot{A}_1^{(\l)})+\ddXl \big)
   \NN
   
  &&&+\displaystyle\sum_{r=1}^\infty \kappa^{-r}
          \left(\Xl A_{r+2}^{(\l)}+2\l \dXl A_{r+1}^{(\l)}+\ddXl A_r^{(l)} + 2\l \Xl \dot{A}_{r+1}^{(\l)}+2\dXl\dot{A}_r^{(\l)}+\Xl\ddot{A}_r^{(\l)}\right)
      ~~~~\Big\}

 \earr
 and
 \barr{rclll}
   (\kappa^2 +\kappa \alpha + \beta)\, d^{(\l)}
   &=&\mb{e}^{\l\,\kappa t}&\Xl\Big\{
   \kappa^2  + \kappa \big( A_1^{(\l)}+  \alpha\big) +  \beta +A_2^{(\l)}
   +\displaystyle\sum_{r=1}^\infty \kappa^{-r}
          \left(A_{r+2}^{(\l)} + \alpha A_{r+1}^{(\l)}+\beta A_r^{(\l)}\right)
      ~~~~\Big\} \nonumber
\earr
Now we compare coefficients for every order of $\kappa$.
\barr{lrrcl}
  \MC{O}(\kappa^2)&:& 0&=&0
  \\
  \MC{O}(\kappa^1)&:& 2\l\dXl&=&\alpha\Xl
  \\
  \MC{O}(\kappa^0)&:& \ddXl+2\l\dot{A}_1^{(\l)}\Xl&=&\beta\Xl
  \\
  \MC{O}(\kappa^{-r})&:& \ddXl A_r^{(\l)}+2\l\dot{A}_{r+1}^{(\l)}\Xl+2\dXl\dot{A}_r^{(\l)}+\Xl\ddot{A}_r^{(\l)}&=&\beta\Xl A_r^{(\l)}
\earr
These equations can be integrated. From $\MC{O}(\kappa^1)$ we find\footnote{Here and in what follows we will always integrate from $0$ to $t$ and insert the appropriate integration constants when necessary.}
\be\label{X_0^l}
   \Xl(t)=\Xl_0~\mb{e}^{\frac \l 2 \int_0^t \alpha(s)ds}~~~\text{with~}\Xl_0 \text{an integration constant~}.
\ee
yielding the following identities
\be\label{Xl identities}
   \frac{\dXl}{\Xl} =\frac{\l}{2}\,\alpha ~~~~~~~\text{and}~~~~~~~
   \frac{\ddXl}{\Xl}= \frac{\l}{2}\,\dot\alpha +\frac{1}{4}\,\alpha^2~~.
\ee
We can use (\ref{Xl identities}) in order to rewrite the $\MC{O}(\kappa^{0})$ and $ \MC{O}(\kappa^{-r})$-expressions of (\ref{equating coefficients of kappa}) to get
\ba\label{dot A^l_1}
   \dot{A}_{1}^{(\l)}
   &=&\frac{\l}{2}\left(\beta -\frac{\l}{2}\,\dot{\alpha} -\frac{1}{4}\,\alpha^2\right)
   \\
   \label{dot A^l_n}
   \dot{A}_{r+1}^{(\l)}
   &=&\frac{\l}{2}\left(\beta -\frac{\l}{2}\,\dot{\alpha} -\frac{1}{4}\,\alpha^2\right)A_r^{(\l)}
    -\frac{1}{2}\, \alpha\, \dot{A}_r^{(\l)}
    -\frac{\l}{2}\, \ddot{A}_r^{(\l)}
\ea
which can be integrated for $r\ge 0$ to
\ba\label{A^l_1}
 A_{1}^{(\l)}(t)
   &=&C_{1}^{(\l)}+
   \frac{\l}{2}\int_0^t \left(\beta(s) 
                                -\frac{\l}{2}\,\dot{\alpha}(s)
                                -\frac{1}{4}\,\alpha^2(s)\right)
                  \,ds 
\\
\label{A^l_n}
   A_{r+1}^{(\l)}(t)
   &=&C^{(\l)}_{r+1}
    -\frac{1}{2}\, \alpha(t)\, A_r^{(\l)}(t)
    -\frac{\l}{2}\, \dot{A}_r^{(\l)}(t)
    +\frac{\l}{2}\int_0^t \left(\beta(s)
                            +\frac{\l}{2}\,\dot{\alpha}(s)
                            -\frac{1}{4}\,\alpha^2(s)\right)A_r^{(\l)}(s)\,ds 
\ea
where $C^{(\l)}_{1}, C^{(\l)}_{r+1}$ are integration constants. Here we have used the fact that $\l^2=1$. Notice the different signs in front of the $\dot{\alpha}$ term in the integrals on the right hand side of 
    (\ref{A^l_1}), (\ref{A^l_n}), which result from a partial integration of the $\alpha\dot{A}^{(\l)}_r$-term in (\ref{dot A^l_n}).
~\\~\\   
The integration constants $C^{(\l)}_{1}, C^{(\l)}_{r+1}$ and  $X_0^{(l)}$ can be fixed by imposing the initial conditions (\ref{ODE 2nd order a initial conditions}) and \linebreak using the Ansatz (\ref{Ansatz general solution for a}).

\subsection{\label{Power-Series Solution to Holonomy ODE}Power-Series Solution to Holonomy ODE for General Edges}
We can now construct a formal solution to the original ODE (\ref{ODE 2nd order a initial conditions}) and to the original ODE system (\ref{first order holonomy ODE system}). To summarize, we have
\ba\label{ansatz a}
   a(\kappa,t)
   &=&\sqrt{m(t)}~\big(A_{(+)}\, d^{(+)}(\kappa,t)+A_{(-)}\,d^{(-)}(\kappa,t)\big)
   \\
   \dot{a}(\kappa,t)
   &=&\frac{1}{2} \,M(t) \,a(\kappa,t) 
   + \sqrt{m(t)}~\big(A_{(+)}\, \dot{d}^{(+)}(\kappa,t)+A_{(-)}\,\dot{d}^{(-)}(\kappa,t)\big)
   ~~.
\ea
with initial conditions
\be
a(\kappa,0)=1 ~~~~~~~~~\dot{a}(\kappa,0)=\kappa~n(0)
\ee
and respectively 
\ba\label{ansatz b}
   b(\kappa,t)
   &=&\sqrt{m(t)}~\big(B_{(+)}\, d^{(+)}(\kappa,t)+B_{(-)}\,d^{(-)}(\kappa,t)\big)
\ea
with initial conditions
\be\label{AB b}
    b(\kappa,0)=0 ~~~~~~~~~\dot{b}(\kappa,0)=\kappa~m(0)
\ee
and $d^{(\pm)}(\kappa,t)$ given by (\ref{LG-Ansatz}).
\paragraph{Fixing $X_0^{(\l)}$.} Without loss of generality we can set the integration constant $X_0^{(\l)}$ in (\ref{X_0^l}) to $X_0^{(\l)}=1$, because as a constant it can be absorbed into $A_{(+)}, A_{(-)}$ respectively $B_{(+)}, B_{(-)}$ in (\ref{ansatz a}) and (\ref{ansatz b}).  
Therefore we have
\be\label{d^l final}
   d^{(\l)}(\kappa,t)
   =\mb{e}^{\l\left(\,\kappa t+\frac 1 2 \int_0^t \alpha(s)ds\right)}
              \left(\sum_{r=0}^\infty\frac{A_r^{(\l)}(t)}{\kappa^r}\right)~~.
\ee
\subsubsection{Fixing integration constants and solution for $b(\kappa,t)$}

We have
\be\label{AB von b}
   b(\kappa,0)
   =0
   =\sqrt{m(0)}\left( B_{(+)}\sum_{n=0}^\infty \frac{A_n^{(+)}(0)}{\kappa^n}
                     +B_{(-)}\sum_{n=0}^\infty \frac{A_n^{(-)}(0)}{\kappa^n}\right)
\ee
Now we compare coefficients for every order of $\kappa$.
\barr{lrrclcl}\label{AB von b 2}
  \MC{O}(\kappa^0)&:& \text{$A_0^{(\l)}=1$, hence ~~~}0&=&B_{(+)}+B_{(-)}
  &\leadsto& B_{(+)}=-B_{(-)}
  \\
  \MC{O}(\kappa^{-{n}})|_{n>0}&:& \text{With $B_{(+)}=-B_{(-)}$ we have ~~~}0&=&A_n^{(+)}(0)-A_n^{(-)}(0)
  &\leadsto& A_n^{(+)}(0)=A_n^{(-)}(0)
\earr
Additionally we have
\ba\label{AB von b'}
   \dot b(\kappa,0)
   =\kappa\,m(0)
   &=&\frac{M(0)}{2}\,b(\kappa,0) +\sqrt{m(0)}\,\left(B_{(+)}\,\dot d^{(+)}(\kappa,0) + 
                                                    B_{(-)}\,\dot d^{(-)}(\kappa,0)\right)
   \NN
   &\stackrel{(\ref{AB b})\atop (\ref{AB von b 2})}{=}&
   \sqrt{m(0)} \,B_{(+)} \left(\dot d^{(+)}(\kappa,0) -\dot d^{(-)}(\kappa,0)\right)
   \NN
   &=& \sqrt{m(0)} \, B_{(+)}
       \left\{\Big( 2\kappa+\alpha(0)\Big)\sum_{n=0}^\infty
              \frac{A_n^{(+)}(0)}{\kappa^n}
              +\sum_{n=0}^\infty \frac{\dot A_n^{(+)}(0)}{\kappa^n}
              -\sum_{n=0}^\infty \frac{\dot A_n^{(-)}(0)}{\kappa^n}
        \right\}
\ea
Comparison of coefficients gives
\barr{lrrclcl}
  \MC{O}(\kappa^1)&:& \text{$A_0^{(\l)}=1$, hence ~~~} 
         m(0)&=&2\sqrt{m(0)}\,B_{(+)}
         &\leadsto& \fbox{$B_{(+)}=-B_{(-)}=\frac{\sqrt{m(0)}}{2}$}
  \\
  \MC{O}(\kappa^0)&:& 0&=&\alpha(0) + 2\,A_1^{(+)}(0)
                       &\leadsto& \fbox{$A_1^{(+)}(0)=A_1^{(-)}(0)=-\frac{\alpha(0)}{2}$}
  \\
  \MC{O}(\kappa^{-n})|_{n>0}&:& \text{$A_n^{(+)}=A_n^{(-)}$} 
  &\leadsto&\multicolumn{3}{l}{\fbox{$A_{n+1}^{(\l)}(0)=-\frac{\alpha(0)}{2}\,A_{n}^{(\l)}(0)
                        -\frac{1}{2}\dot A_{n}^{(+)}(0) 
                        +\frac{1}{2} \dot A_{n}^{(-)}(0)$}}
\earr
If we compare this to (\ref{A^l_1}), (\ref{dot A^l_n})  we get
\be
   C_1^{(+)}=C_1^{(-)}=-\frac{\alpha(0)}{2}
   ~~~~\text{and}~~~~
   C_{n+1}^{(+)}=\frac{1}{2}\dot A_n^{(-)}(0)~~,~~
   C_{n+1}^{(-)}=-\frac{1}{2}\dot A_n^{(+)}(0)~~.~~
\ee

\subsubsection{Solution for $a(\kappa,t)$} 
We have

\be\label{AB von a}
   a(\kappa,0)
   =1
   =\sqrt{m(0)}\left( A_{(+)}\sum_{n=0}^\infty \frac{A_n^{(+)}(0)}{\kappa^n}
                     +A_{(-)}\sum_{n=0}^\infty \frac{A_n^{(-)}(0)}{\kappa^n}\right)
\ee
Now we compare coefficients for every order of $\kappa$.
\barr{lrrclcl}\label{AB von a 2}
  \MC{O}(\kappa^0)&:& \text{$A_0^{(\l)}=1$, hence~~~}1&=&\sqrt{m(0)}\big(A_{(+)}+A_{(-)}\big)
  &\leadsto&\fbox{$ A_{(-)}=[m(0)]^{-\frac{1}{2}}-A_{(+)}$}
  \\
  \MC{O}(\kappa^{-{n}})|_{n>0}&:& 0&=& A_{(+)}A_n^{(+)}(0)
                                      +A_{(-)}A_n^{(-)}(0)
  &\leadsto& \fbox{$A_n^{(-)}(0)=-\frac{A_{(+)}}{A_{(-)}}\,A_n^{(+)}(0)$}
\earr
Additionally we have
\ba\label{AB von a'}
   \dot a(\kappa,0)
   =\kappa\,m(0)
   &=&\frac{M(0)}{2}\,a(\kappa,0) +\sqrt{m(0)}\,\left(A_{(+)}\,\dot d^{(+)}(\kappa,0) + 
                                                    A_{(-)}\,\dot d^{(-)}(\kappa,0)\right)
   \NN
   &\stackrel{(\ref{AB von a})\atop (\ref{AB von a 2})}{=}&
   \frac{M(0)}{2}+
   \sqrt{m(0)} \left\{ 
                   A_{(+)}\left[
                       \left(\kappa+\frac{\alpha(0)}{2} \right) 
                       \sum_{n=0}^\infty \frac{A_n^{(+)}(0)}{\kappa^n} 
                      +\sum_{n=0}^\infty \frac{\dot A_n^{(+)}(0)}{\kappa^n} 
                   \right] \right.
\NN&&                   
                   ~~~~~~~~~~~~~~~~~~~~+
                  \left. 
                  A_{(-)}\left[
                       -\left(\kappa+\frac{\alpha(0)}{2} \right) 
                       \sum_{n=0}^\infty \frac{A_n^{(-)}(0)}{\kappa^n} 
                      +\sum_{n=0}^\infty \frac{\dot A_n^{(-)}(0)}{\kappa^n} 
                   \right] 
               \right\}
\ea
Comparison of coefficients gives ($A_0^{(\l)}=1$)
\barr{lrrclcl}
  \MC{O}(\kappa^1)&:& 
         n(0)&=&\sqrt{m(0)}\,\big(A_{(+)}-A_{(-)}\big)
         &\stackrel{(\ref{AB von a 2})}{\leadsto}& \fbox{$A_{(\pm)}=\frac{1\pm n(0)}{2\sqrt{m(0)}}$}
  \\
  \MC{O}(\kappa^0)&:& 0&=&\multicolumn{3}{l}{
  \frac{M(0)}{2}
  +\sqrt{m(0)}\Big\{
                A_{(+)}\big(A_1^{(+)}(0)+\frac{\alpha(0)}{2}\big)
               -A_{(-)}\big(A_1^{(-)}(0)+\frac{\alpha(0)}{2}\big)              
              \Big\}
                       }
  \\
  &&&&&\leadsto&\fbox{$A_1^{(\pm)}(0)=-\frac{\alpha(0)\,n(0)+M(0)}{2(1\pm n(0))}$}
  \\
  \MC{O}(\kappa^{-n})|_{n>0}&:& 
  0&=&\multicolumn{3}{l}{
       A_{(+)}\Big[A_{n+1}^{(+)}(0)+\frac{\alpha(0)}{2}A_n^{(+)}(0) + \dot A_n^{(+)}(0) \Big]
      -A_{(-)}\Big[A_{n+1}^{(-)}(0)+\frac{\alpha(0)}{2}A_n^{(-)}(0) - \dot A_n^{(-)}(0) \Big]
      }  
      \\
  &&&&&\stackrel{(\ref{AB von a 2})}{\leadsto}&
  \fbox{$A_{n+1}^{(+)}(0)=-\frac{\alpha(0)}{2}A_n^{(+)}(0)
                          -\frac{1}{2}\dot A_n^{(+)}(0) 
                          -\frac{1}{2}\frac{A_{(-)}}{A_{(+)}}\dot A^{(-)}_n(0)$}
 \\                         
 &&&&&\stackrel{(\ref{AB von a 2})}{\leadsto}&
  \fbox{$A_{n+1}^{(-)}(0)=-\frac{\alpha(0)}{2}A_n^{(-)}(0)
                          +\frac{1}{2}\dot A_n^{(-)}(0) 
                          +\frac{1}{2}\frac{A_{(+)}}{A_{(-)}}\dot A^{(+)}_n(0)$}                         
  
\earr
where we have just multiplied the second to last row by $-\frac{A_{(+)}}{A_{(-)}}$ in order to arrive at the last row. If we compare this to (\ref{A^l_1}), (\ref{dot A^l_n})  for $t=0$ we get
\ba
   C_1^{(+)}=-\frac{\alpha(0)\,n(0)+M(0)}{2\big(n(0)+1\big)}
   &~~~~\text{and}~&
   C_{n+1}^{(+)}=-\frac{1}{2}\,\frac{A_{(-)}}{A_{(+)}}\dot A_n^{(-)}(0)
                =~~\, \frac{1}{2}\,\frac{n(0)-1}{n(0)+1}\dot A_n^{(-)}(0)
   \NN
   C_1^{(-)}=-\frac{\alpha(0)\,n(0)+M(0)}{2\big(n(0)-1\big)}
   &&C_{n+1}^{(-)}=~~\,\frac{1}{2}\,\frac{A_{(+)}}{A_{(-)}}\dot A_n^{(+)}(0)
                = -\frac{1}{2}\,\frac{n(0)+1}{n(0)-1}\dot A_n^{(+)}(0)
\ea

\subsubsection{\label{Final Solution}Final Solution}

We have thus constructed a formal solution to the ODE-system (\ref{first order holonomy ODE system}) with initial conditions (\ref{ODE 2nd order a initial conditions})

\ba
   a(\kappa,t)
   &=&\sqrt{m(t)}~\big(A_{(+)}\, d^{(+)}(\kappa,t)+A_{(-)}\,d^{(-)}(\kappa,t)\big)
\NN
  b(\kappa,t)
   &=&\sqrt{m(t)}~\big(B_{(+)}\, d^{(+)}(\kappa,t)+B_{(-)}\,d^{(-)}(\kappa,t)\big)~~.
\ea
Using arc length parametrization $\ol{m}(t)\,m(t)+n(t)^2=1$ and 
\[
   \alpha(t):=\dot{n}(t)-M(t)\,n(t)~~~~~~~~
   \beta(t):=\frac{1}{4}M^2(t)-\frac{1}{2}\dot{M}(t)
\]
we then have for $\l=\pm 1$
\be\label{final A definition}
   d^{(\l)}(\kappa,t)
   =\mb{e}^{\l\left(\,\kappa t+\frac 1 2 \int_0^t \alpha(s)ds\right)}
              \left(\sum_{r=0}^\infty\frac{A_r^{(\l)}(t)}{\kappa^r}\right)~~.          
\ee
with 
\ba\label{final A recurrence}
 A_{0}^{(\l)}(t)&=&1
 \NN
 A_{1}^{(\l)}(t)
   &=&C_{1}^{(\l)}+
   \frac{\l}{2}\int_0^t \left(\beta(s) 
                                -\frac{\l}{2}\,\dot{\alpha}(s)
                                -\frac{1}{4}\,\alpha^2(s)\right)
                   A_r^{(\l)}(s)\,ds 
\NN
   A_{r+1}^{(\l)}(t)
   &=&C^{(\l)}_{n+1}
    -\frac{1}{2}\, \alpha(t)\, A_r^{(\l)}(t)
    -\frac{\l}{2}\, \dot{A}_r^{(\l)}(t)
    +\frac{\l}{2}\int_0^t \left(\beta(s)
                            +\frac{\l}{2}\,\dot{\alpha}(s)
                            -\frac{1}{4}\,\alpha^2(s)\right)A_r^{(\l)}(s)\,ds~~~, 
\ea
where the last line is valid for $r\ge 1$.
For $a(\kappa,t)$ we have the integration constants
\be\nonumber     
    A_{(\l)}=\l\cdot \frac{n(0)+\l}{2\sqrt{m(0)}}
    ~~~~
    C_1^{(\l)}=-\frac{\alpha(0)\,n(0)+M(0)}{2\big(n(0)+\l\big)}
    ~~~~
    C_{r+1}^{(\l)}=\frac{\l}{2}\cdot\,\frac{n(0)-\l}{n(0)+\l}\dot A_r^{(-\l)}(0)
\ee
and for $b(\kappa,t)$ we have the integration constants
\be\nonumber     
    B_{(\l)}=\l\cdot \frac{\sqrt{m(0)}}{2}
    ~~~~
    C_1^{(\l)}=-\frac{\alpha(0)}{2}
    ~~~~
    C_{r+1}^{(\l)}=\frac{\l}{2}\cdot \dot B_r^{(-\l)}(0)~~~.
\ee

Given these solutions in terms of a formal power series it remains to discuss the finiteness properties of this series. This issue will be discussed in the next section.

\section{\label{Asymptotics}Asymptotics and Finiteness of the Solution}
\subsection{Asymptotic Behavior of $SU(2)$ Holonomies}

Using the Liouville-Green method and Horn asymptotics (see e.g. \cite{Olver}) it is straight forward to find the asymptotic behavior of the holonomy differential equation. We present the general procedure here and refer to appendix \ref{Appendix epsilon schranke} for details.

Recall that the general second order linear ordinary differential equation $\ddot a(t)=A(t) \dot a(t)+ B(t) a(t)$ is transformed into standard form $\ddot d=(B(t)+(\frac 1 4 A^2(t)-\frac 1 2 \dot A(t)))d(t)$ using $a(t)=d(t)\exp(\frac 1 2 \int_{t_o}^t ds A(s))$. This transforms the holonomy differential equation (in arc-length parametrization for $t$) into
\begin{equation}
 \ddot d(t) = \left(\kappa^2 + \kappa \alpha(t)+\beta(t)\right) d(t),
\end{equation}
where we introduced $\kappa:=\i c$. The Liouville-Green Ansatz for the solution is
\begin{equation}
 d^{(\pm)}(t)=\exp\left(\pm\kappa t\pm \frac 1 2 \int_o^t ds~\alpha(s)\right)\left(1+\sum_{n=1}^\infty\frac{A^{(\pm)}_n(t)}{\kappa^n}\right),
\end{equation}
which suggests that $d^{(\pm)}(t)\to C^{(\pm)} e^{\kappa t}$ as $|\kappa|\to \infty$, with prefactor $C^{(\pm)}$. To prove this behavior, one needs convergence of this Ansatz, which is rarely given. In fact the Liouville-Green Ansatz for a generic holonomy diverges. However, any truncated Ansatz $\sum_{k=1}^n\frac{A^{(\pm)}_k(t)}{\kappa^k}$ is well behaved (if the $n$-th derivatives of $\alpha,\beta$ are bounded) and hence we consider the truncation error
\begin{equation}
 \epsilon^{(\pm)}_n(t)=d^{(\pm)}(t)-\sum_{k=1}^n\frac{A^{(\pm)}_k(t)}{\kappa^k}.
\end{equation}
The asymptotic behavior can the be obtained from bounding the truncation error using the following steps
\begin{enumerate}
\item Insert the truncation error into the ODE to obtain an equation for $\epsilon^\pm_n(t)$. This ODE can be solved using variation of constants in Horn's asymptotic Ansatz $Z^\pm=\left(\kappa+\frac 1 2\alpha\right)^{-\frac 1 2}\exp\left(\pm(\kappa t+\frac 1 2 \int \alpha)\right)$. 

\item The key of Horn's Ansatz is that it solves the ODE to all positive powers of $\kappa$, hence the kernel of the integral equation obtained form the variation of constants does not contain any positive powers of $\kappa$ and thus yields finite bounds on the solution when $|\kappa|\to\infty$.

\item Finiteness of the coefficients of the truncated Liouville-Green series together with the bound on the truncation error yields the anticipated asymptotic behavior of  $d^{(\pm)}(t)\to C^{(\pm)} e^{\pm\kappa t}$ as $|\kappa|\to \infty$ and thus also for $a(t)=d(t)\exp(\frac 1 2 \int_{t_o}^t ds A(s))$, since $A$ is independent of $\kappa$.

\item The asymptotic behavior of the matrix elements of the holonomy is then determined by the asymptotic behavior of the linear combination of $d^{(\pm)}(t)$ that satisfies the initial condition for the specific matrix element, which yields the anticipated asymptotic periodicity of the holonomy. 
\end{enumerate}
A significant simplification to estimate of the truncation error can be obtained by using methods developed in \cite{Dunster1993}. There it is assumed that the edges along which the holonomy is integrated are holomorphic\footnote{In this case one can apply Cauchy's integral formula in order to replace derivatives of $\alpha(t), \beta(t)$ in (\ref{final A recurrence})  by integral expressions.} in a finite radius in the complex plane around the part of the real $t$-axes that is integrated, because this yields a very simple bound on the Liouville-Green coefficients $d_n^{(\pm)}(t)$. This method is applicable if one is able to satisfy initial conditions s.t. all integration constants $C_r^{(\l)}$ in the Liouville Green recursion (\ref{final A recurrence}) vanish. Unfortunately this requirement is in general not compatible with the initial conditions (\ref{ODE 2nd order a initial conditions}). A solution to (\ref{first order holonomy ODE system}) fulfilling the initial condition  (\ref{ODE 2nd order a initial conditions}) does in general not have vanishing $C_r^{(\l)}$. This is only possible in very special cases\footnote{See the paragraph on Convergence at the end of this section.}.\\

Let us now discuss this result: By construction the series (\ref{final A definition})  provides a formal solution to the ODE (\ref{LG-Vorbereitung 2}). However it is not not obvious whether this series is finite nor whether it converges. 

\paragraph{Finiteness.} As explicitly shown in appendix section \ref{Appendix epsilon schranke}, for analytic edges the series (\ref{final A definition}) is finite: at every finite inverse order of $\kappa$ the rest term   
\barr{lcl}
  \epsilon^{(\l)}_{n>0}(\kappa,t)
  &:=&\displaystyle
  d^{(\l)}(\kappa,t)-\mb{e}^{\l\kappa t+\frac{\l}{2}\int_0^t \alpha(s)\,ds}\sum_{k=0}^{n-1}\frac{A^{(\l)}_k(t)}{\kappa^k}
\earr
obeys the explicit bound   
\barr{lcl}
  \big|\epsilon^{(\l)}_n(\kappa,t)\big|
  &\le& \displaystyle
  \frac{\exp\left(\int_0 ^t |\alpha(s)|\,ds\right)}{\lambda(\kappa,t)}
  \left(\int_0^t \big|\Delta^{(\l)}_n(\kappa,\tau)\big|\,d\tau\right)
           \exp\left(\exp\left(\frac 1 2\int_0^t |\alpha(s)|ds\right) \frac{\int_0^t |\chi(\kappa,s)|\,ds}{\lambda(\kappa,t)}\right)\,
  \NN
  &\le&  \displaystyle  
  \frac{2~\Phi(t)^2~ \int_0^t \Phi(\tau)~\big|\dot A^{(\l)}_{n}(\tau)\big|\,d\tau}{\lambda(\kappa,t)~|\kappa|^{n-1}}
       ~~ \exp\left( \frac{\Phi(t)\int_0^t |\chi(\kappa,s)|\,ds}{\lambda(\kappa,t)}\right)\, 
\earr
where $\lambda(\kappa,t):=\inf_{0\le\tau\le t}\{|\kappa+\frac 1 2 \alpha(\tau)|\}$ and $|\Delta^{(\l)}_n(\kappa,t)|
 =|\mb{e}^{\l \kappa t}\Xl(t) \frac{ 2\l \dot A^{(\l)}_{n}(t)}{\kappa^{n-1}} |
 \le |\kappa|^{n-1}~\exp\left(\int_0 ^t |\alpha(s)|\,ds\right)~|\dot A^{(\l)}_{n}(t)|$. 
Here we have set  $X^{(\l)}(t)=\mb{e}^{\frac \l 2 \int_0^t \alpha(s)ds}$ according to section to section \ref{LG}. The quantity $\chi(\kappa,s)$ is given by 
$\chi(\kappa,s)=\beta(s)
                    -\frac{\alpha^2(s)}{4}
                    -\frac{3\dot \alpha^2(s)}{16(\kappa+ \frac 1 2 \alpha(s))^2}
                    +\frac{\ddot \alpha(s)}{4(\kappa+\frac 1 2 \alpha(s))}$
and the functions $\alpha(t), \beta(t)$ are given as in (\ref{LG-Vorbereitung}). In the last line we have introduced the shorthand $\Phi(t):=\exp\left(\frac{1}{2}\int_0 ^t |\alpha(s)|\,ds\right)$. Hence for fixed $t$ we find that  $\big|\epsilon^{(\l)}_n(\kappa,t)\big|\sim\mathcal{O}(\kappa^{-n})$. At the same time, given the finiteness of the interval $[0,t]\subset\mb{R}$ and assuming the analyticity properties of the functions $\alpha(t), \beta(t)$ as in section \ref{RAL}, it is obvious from (\ref{final A recurrence}) that for every $r<\infty$ it holds that $\big|A_{r+1}^{(\l)}(t)\big|$ is bounded from above, because it is a finite combination of lower order terms.

\paragraph{Convergence.} In section \ref{Estimate of the Error Term:Dunster} a general estimate is given for $\big|A_{r+1}^{(\l)}(t)\big|$ is derived which only depends on $r$. There the results of \cite{Dunster1993} are applied. These results only hold under quite restrictive assumptions on the recursion relation (\ref{final A recurrence}), in particular it requires the integration constants $C_{n}^{(\l)}$ to be identical zero. Moreover the functions $\alpha(t),\beta(t)$ are required to be holomorphic in a small strip around the $\mb{R}$-axis in the complex plane, in order to use Cauchy's integral formula to replace derivatives with respect to $t$ by contour integrals. Unfortunately these requirements are too restrictive to be applied for the $\big|A_{r+1}^{(\l)}(t)\big|$ for general curves $\gamma$. Interestingly the requirement $C_{n}^{(\l)}=0~~\forall n>1$ is fulfilled for spiral arcs, for which however the general explicit solution is known \cite{Brunnemann:2007du}.

\subsection{Asymptotic Behavior of Holonomy Matrix Elements for Symmetric Connections}

In summary, the last subsection has shown:
\begin{lemma}\label{Lemma1}
  Given the analyticity condition (\ref{equ:analyticity-condition}), (\ref{generalized t_k}) and quasi-arc-length parametrization \linebreak  $n^2(t)+|m(t)|^2=1$ there exists a $t_2$ s.t. the $SU(2)$ holonomy matrix elements satisfy for $t<t_2$:
  \begin{equation}\label{asymptotic ME}
    \begin{array}{rcl}
      a(\kappa,t)&\to&\displaystyle
      \sqrt{\frac{m(t)}{m(0)}}
      \left\{ 
       \cosh\left(\kappa t+\frac 1 2 \int_0^t  \alpha(s)\,ds\right)-\mb{i}
      n(0) \, \sinh\left(\kappa t+\frac 1 2 \int_0^t  \alpha(s)\,ds\right)
      \right\}\\
      b(\kappa,t)&\to&\displaystyle
      -\mb{i}\sqrt{m(t)\,m(0)}\,
      \sinh\left(\kappa t+\frac 1 2 \int_0^t  \alpha(s)\,ds\right)
    \end{array}
  \end{equation}
  in the limit $|c|\to \infty$.
\end{lemma}
Notice that this gives the asymptotics only. In order to re-derive the property that $|a(\kappa,t|^2+|b(\kappa,t)|^2=1$ one has to start from the general solutions of section \ref{Final Solution}, compute  $|a(\kappa,t)|^2$ and $|b(\kappa,t)|^2$ and then take the limit\footnote{There are mixing terms which contribute then. }. 
 
An edge $e$ is parametrized by $0\le t \le T$, which allows us to define the inverse edge $e^{-1}(t):=e(T-t)$. Lemma \ref{Lemma1} thus implies that there exists $t_2(e)>0$ and $t_2(e^{-1})>0$, where the asymptotics (\ref{asymptotic ME}) holds. We can thus restrict our attention to the compact interval $t_2(e)/2 \le t \le T-t_2(e^{-1})/2$. For any point $t$ in this interval we are able to onsider $e^+_t(\lambda):=e(t+\lambda)$ and $e^-_t(\lambda):=e(t-\lambda)$, and lemma \ref{Lemma1} states that there are $t_2^+,t_2^->0$ such that the asymptotics (\ref{asymptotic ME}) holds for $0\le t \le t_2^+$ on $e^+_t$ and analogously for $e^-_t$, implying that any point $t$ in the interval $t_2(e)/2 \le t \le T-t_2(e^{-1})/2$ has an open neighborhood where the asymptotics holds. Hence by compactness of the interval $t_2(e)/2 \le t \le T-t_2(e^{-1})/2$ one can find a finite open covering of this interval where the asymptotics holds for each element of the covering. Thus the asymptotics of the holonomy is given by a finite number of matrix products the matrix elements of each having the asymptotics (\ref{asymptotic ME}), which imples that the holonomy of the entire edge has exponential asymptotics in $\kappa$.

Now we introduce:
\begin{definition}
  A complex-valued function is called asymptotically almost periodic, iff it can be written as the sum of a continuous almost periodic function and a continuous function that vanishes at infinity and at zero.
\end{definition}
\begin{lemma}\label{lem:AAP-algebra}
  The continuous asymptotically almost periodic functions form an algebra over $\mathbb C$, and each element splits uniquely into the almost periodic functions plus functions vanishing at infinity. 
\end{lemma}
\paragraph{\it proof:} Let $f_i$ be asymptotically almost periodic, so there exist continuous almost periodic functions $a_i$ and continuous functions $b_i$ vanishing at infinity, such that $f_i=a_i+b_i$. \\~\\
{\it algebra:}~~ $\alpha f$ is asymptotically almost periodic, since for $\alpha \in \mathbb C$: $\alpha a$ is almost periodic and $\alpha b$ vanishes at infinity. $f_1+f_1$ is asymptotically almost periodic since $a_1+a_2$ is almost periodic and $b_1+b_2$ vanishes at infinity. $f_1 f_2$ is asymptotically almost periodic, since $a_1a_2$ is almost periodic and $a_1b_2+a_2b_1+b_1b_2$ vanishes at infinity.\\~\\
{\it uniqueness:~~} Assume $f$ is asymptotically almost periodic and there exists continuous $a_1\ne a_2$ almost periodic and continuous $b_1\ne b_2$ s.t. $f=a_1+b_1=a_2+b_2$, so $a_1-a_2=b_2-b_1$, which implies that $a_1-a_2$ vanishes at infinity, hence $a_1=a_2$, thus $f\mapsto a_1+b_1$ is unique. If $c=0$ the original ODE system (\ref{first order holonomy ODE system}) has constant solutions which are almost periodic. 
 $\square$\\~\\
Let us recall that a gauge-variant spin-network function $T_\gamma$ is a finite collection of edges $\gamma=(e_1,...,e_n)$ with a matrix-element of an irreducible representation $\rho$ of the gauge group ($SU(2)$ in the present case) associated with each edge, such that $T_\gamma(A)=\prod_{i=1}^n \rho^{j_i}(h_{e_i}(A))_{m_in_i}$. Using that a matrix element of an irreducible representation is a polynomial function of an the matrix elements of the fundamental representation, we find using lemma \ref{lem:AAP-algebra}:
\begin{Theorem}Let $c A_*$ denote the symmetric connection as described in section \ref{Holonomy ODE for Homogeneous Isotropic Cosmological Model}.
  Given a spin network function $T_\gamma$ s.t. each edge $e_i\in \gamma$ can be written $e_i=e_{i,1}\circ...\circ e_{i,k_i}$ such that each $e_{i,j}$ satisfies the analyticity condition (\ref{equ:analyticity-condition}), then $T_\gamma(c A_*)$ is an asymptotically almost periodic function of $c$.
\end{Theorem}

This brings up the idea, to group the set of all spin network functions $T_\gamma$ into equivalence classes, where  $T_\gamma\sim T'_{\gamma'}$ are called equivalent if $T_\gamma(A)|_{cA_*}= T'_{\gamma'}(A)|_{cA_*}$, that is if they coincide when evaluated on a symmetric connection $cA_*$. In section \ref{ExplQuantSymm} this observation will be used in order to construct an embedding of the extended configuration space of loop quantum cosmology (given by the set  $AAP(\mb{R}$ of asymptotically almost periodic functions) into $Cyl(\ol{\MC{A}})$, the configuration space of full loop quantum gravity.
  
\subsection{\label{Generalization to the Full Configuration Space}Generalization to the Full Configuration Space}

Remarkably, the previous construction generalizes straightforwardly to {\it any}\footnote{Not necessarily symmetric.} bounded reference connection $A_*$: let us consider a one-dimensional subspace of $\ol{\MC{A}}$, the space of generalized connections. Let this subspace contain connections of the form $c A_*$, where $A_*$ is a bounded $su(2)$-valued reference one-form, i.e. its components $(A_*)_a^i$ are bounded in a trivialization $A_*=(A_*)_a^I~ dx^a\otimes \tau_I$ .
Using the notation
\begin{equation}
  \begin{array}{rcl}
    n(t)&=&{(A_*)}^3_{a}(e(t))\dot e^a(t)\\
    m(t)&=&\Big({(A_*)}_{a}^1(e(t))-\i {(A_*)}_{a}^2(e(t)))\Big)\dot e^a(t),
  \end{array}
\end{equation}
we impose the "quasi arc-length" parametrization condition $n^2(t)+|m(t)|^2=1$, such that the holonomy ODE (\ref{Setup 1}) becomes
\begin{equation}
  \ddot w(t)=M(t)\dot w(t)+\left(ic(\dot n(t)-M(t)n(t))-c^2\right)w(t),
\end{equation}
where $M(t)=\frac{\dot m(t)}{m(t)}$ and $w=a$ or $b$ with initial condition $a(0)=1$, $b(0)=0$. We thus have a tool to investigate the compactification of the configuration space of LQG in a general \lq\lq radial\footnote{We understand radial in the sense of a fictitious Banach-space.} \rq\rq through the induced Gel'fand topology on the radial coordinate $c$. To re-phrase this: {\it every} gauge variant spin-network function $T_\gamma$ exhibits an asymptotically almost periodic dependence on $c$ if the connection $A$ is parametrized\footnote{For a general connection $A\in\ol{\MC{A}}$ with bounded components $A_a^I(x)<\infty~~\forall x\in\Sigma$ and $\forall a,I=1,2,3$  in a chosen trivialization, choose e.g. $c:=\sup_{x,a,I}|A_a^I(x)|$ and introduce $(A_*)_a^I(x):=c^{-1} A_a^I(x)$. } as $cA_*$.

\section{\label{SymmQuantCon}Quantum Symmetry Reduction}
\subsection{Symmetric Quantum Connections}
A classical connection $\omega$ (on a trivial bundle) is determined by the connection one-form $A=\sigma^* \omega$ obtained as its pullback under a section $\sigma$. A connection one-form then defines a homeomorphism from the groupoid $\mathcal P$ of piecewise analytic paths to the gauge group $\mathbb G$ by
\barr{rcl}
 A: \mathcal P &\to& G ~~,~~~  e \mapsto h_e(A),
\earr
where $h_e$ denotes the holonomy along the path $e$. A quantum connection is thus conveniently defined as an element of $Hom(\mathcal P,\mathbb G)$, which imposes no further restriction on the homeomorphism. This is in contrast to a classical connection where differentiability allows to reconstruct the connection components unambiguously form a much smaller subgroupoid $\mathcal P_o$, that is local and can be constructed using in each chart $(U^d,\{\phi^a\}_{a=1}^d)$ as follows: 

The base space is $\mathcal P_o^{(o)}=U$. At each point $x_o\in U^d$ we consider the $d$ linearly independent coordinate directions $d\phi^a$ and the $d$ corresponding families of integral curves $e^a_{x_o,t}:(0,t)\to  U^d$ which we consider as arc-length parametrized in the Euclidean metric on $\mathbb R^d$. The elements of the groupoid $\mathcal P_o$ then consist of all finite concatenations of these integral curves modulus zero paths and the source map is $s(e^a_{x_o,t})=x_o$ and range map is $r(e^a_{x_o,t})=e^a_{x_o,t}(t)$. The components of a classical connection are then by construction completely determined in terms of the restriction to $\mathcal P_o$ through the holonomy differential equation
\begin{equation}\label{equ:connection-reconstruction}
 A^a_I(x_o) \tau^I = \left.\partial_t h_{e^a_{x_o,t}}(A)\right|_{t=0}.
\end{equation}
Let us now consider the subspace $\mathcal S$ of classical connections that are invariant under the action of a symmetry group $\mathbb H$. We can use equation \ref{equ:connection-reconstruction} to define a classical symmetric connection (tautologically) as follows:
\begin{definition}[Classical Symmetry]\label{defi:classical-symmetry}
 A morphism from $A:\mathcal P\to \mathbb G$ is called classically $\mathbb H$-symmetric iff there exists a diffeomorphism $\phi$ and gauge transformation $g$ and an element $B\in \mathcal S$ s.t. $A(e)=g^{-1}(s(e))h_{\phi(e)}(B)g(r(e))$ for all $e\in \mathcal P_o$. 
\end{definition}
Since $\mathcal P_o$ is a subgroupoid of $\mathcal P$, we can immediately apply this definition to quantum connections. However, using this definition does not seem to capture a useful notion of symmetry for quantum connections as can be seen from a simple example: let $B\in \mathcal S$ and $\bar B \notin \mathcal S$ and define the homeomorphism as the extension by groupoid composition of a generating set that contains a generating set of $\mathcal P_o$ and generators that can not be decomposed into elements of $\mathcal P_o$:
\begin{equation}\label{equ:funky-symmetry}
 A: e \mapsto \left\{\begin{array}{rcl} h_e(B)&:&e \in \mathcal P_o\\ h_e(\bar B)&:& e\in \bar{\mathcal P}_o \end{array}\right.
\end{equation}
where $\bar{\mathcal P}_o$ denotes all elements of $\mathcal P$ which do not admit nontrivial decomposition that contains an element of $\mathcal P_o$. This quantum connection does not appear symmetric for almost all paths, except for the ones in $\mathcal P_o$. We are thus lead to use a stronger definition of symmetry for quantum connections:
\begin{definition}[Quantum Symmetry]\label{defi:quantum-symmetry}
 A morphism $A:\mathcal P\to \mathbb G$ is called quantum $\mathbb H$-symmetric iff there exists a diffeomorphism $\phi$ and gauge transformation $g$ and an element $B\in \mathcal S$ s.t. $A(e)=g^{-1}(s(e))h_{\phi(e)}(B)g(r(e))$ for all $e\in \mathcal P$.
\end{definition}
These two definitions of symmetry applied to the morphism defined by the holonomy of a classical connection coincide due to the invertability of \ref{equ:connection-reconstruction}, but weed out strange quantum connections as the one defined through equation \ref{equ:funky-symmetry}. 

Notice that usual Loop Quantum Cosmology is motivated by a model of Loop Quantum Gravity defined on $\mathcal P_o$, so the difference between these two definitions does not occur there and one usually uses definition \ref{defi:classical-symmetry}.

\subsection{\label{ExplQuantSymm}An Explicit Quantum Symmetry Reduction}

The configuration algebra of standard Loop Quantum Cosmology is the closed span of exponential functions of the connection parameter $c$, so the exponential functions are a dense (using the $||.||_\infty$-norm) subset of the configuration operators. The assumed relation $r$ between exponential functions $e^{i c t}$ in standard Loop Quantum Cosmology with full Loop Quantum Gravity configuration operators is given by the observation that holonomy matrix $h_t$ elements along straight edges of length $t$ exhibit this exponential dependence on $c$, so $r(e^{i c t})=h_t$. This relation extends straightforwardly to the span of the exponential functions by linearity. To extend it to the closed span however, one needs to assume a bound $||r(f)||_{LQG} \le ||f||_{\infty}$ to ensure convergence. 

In this first attempt we set difficulties\footnote{It is in particular not obvious whether the completion of the embedding in the full LQG-norm lies within the Hilbert space of full LQG. } concerning issues of convergence and completion aside for now, but caution that one might be forced to amend the definition of $r$ to extend it to the closure. We will denote by $C_c(\mb{R})$ functions of compact support and by $C_o(\mb{R})$ functions, which vanish at infinity. 

The construction strategy for the quantum embedding follows \cite{Koslowski:2006fx}: Let $A(c)$ be the embedding of the symmetry reduced configuration space, parametrized by $c$ into the space of all connections, so $A$ can be used to pull-back any configuration observable (in particular cylindrical functions) $T(A)$ to a function $T(A(c))$ on the symmetry reduced configuration space, which defines equivalence classes of configuration observables in the full theory whose pull-back to the symmetry reduced configuration space coincides. Thus, a symmetry reduction is encoded in these equivalence classes. To relate the symmetry reduced model with the full theory, one constructs an embedding of the symmetry reduced configuration observables $f(c)$. For this we have to find a representative $r(f)$ in the full theory for each equivalence class, such that $\left.r(f)\right|_{A(c)}=f(c)$.

The difference between our present treatment and standard loop quantum cosmology is that our treatment forces us not to restrict ourselves to continuous almost periodic functions $f_a\in AP(\mathbb R)$, but to continuous asymptotically almost periodic functions $f\in AAP(\mathbb R)$. Given a continuous asymptotically almost periodic function $f\in AAP(\mathbb R)$, we can (by definition) find a pair $f_a \in AP(\mathbb R), f_o \in C_o(\mathbb R)$ s.t. $f=f_a+f_o$. Lemma \ref{lem:AAP-algebra} provides that the pair $f_a,f_o$ is unique. 
In order to construct an explicit embedding of $AAP(\mb{R})$ into $Cyl(\ol{\MC{A}})$, the configuration space of full LQG, we are now going to construct a generating set of functions, which allows us to describe every $f\in AAP(\mb{R})$ {\it without} referring to the explicit series  (\ref{final A definition}). For this we start with the definition of some functions. 

\begin{figure}[htbp]
\center
\includegraphics{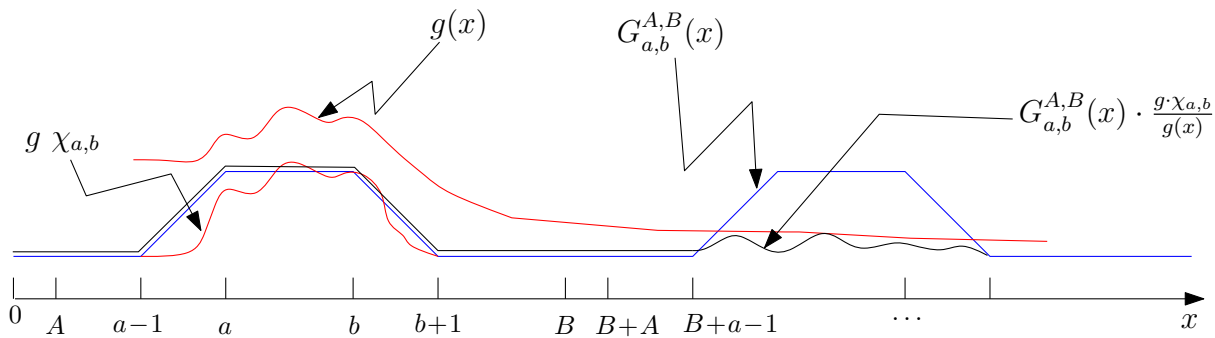}
\caption{Functions(schematically) used in the construction of the embedding.}
\label{embedding}
\end{figure}

For each $a\le b$ let us define $\chi_{a,b}\in C_c(\mathbb R)$ through
\begin{equation}
 \chi_{a,b}(x):=\left\{
                  \begin{array}{rcl}
                    0 &:& x<a-1\\
                    x-a+1&:& a-1\le x< a\\
                    1 &:& a \le x < b\\
                    b-x+1&:& b\le x < b+1\\
                    0 &:& x\ge b+1
                  \end{array}
                \right. .
\end{equation}
For each $a\le b$ and $f\in C_c(\mathbb R)$ with $\text{supp}(f)\subset [a,b]$ we introduce the continuous $(b-a+2)$-periodic function $f_{a,b}\in AP(\mathbb R)$ defined through its values at the defining interval
\begin{equation}
  f_{a,b}(x):=\left\{
                \begin{array}{rcl}
                  0 &:& a-1 < x \le a\\
                  f(x)&:& a < x \le b\\
                  0 &:&b \le x <b+1
                \end{array}
              \right. ,
\end{equation}
and extended to all of $\mb{R}$ by $(b-a+2)$-periodicity\footnote{That is, we just successively concatenate copies of the defining interval along the real axis in order to obtain a periodic function. }.
For a specific positive function $g(x)$ and each $A<a-1,a<b,b+1<B$ we define the continuous $(B-A)$ -periodic function $G^{A,B}_{a,b}\in AP(\mathbb R)$ defined on the initial period $A\le x<B$ as
\begin{equation}
 G^{A,B}_{a,b}(x):=\left\{
                        \begin{array}{rcl}
                          0 &:& A \le x < a-1\\
                          \frac{\chi_{a,b}(x)}{g(x)} &:& a-1 \le x < b+1\\
                          0 &:& b+1 \le x < B
                        \end{array}
                      \right. ,
\end{equation}
and extended to all of $\mathbb R$ by $(B-A)$-periodicity. Let us furthermore assume $g\in C_o(\mathbb R)$ s.t. $\forall x:g(x)>0$ and $\exists g_o>0: g(x)\le \frac{g_o}{|x|^2}$ for $|x|>x_o$. 
\begin{lemma}\label{lem:AAPapproximation}
   Every $f\in AAP(\mathbb R)$ there is a Cauchy sequence of sums of exponential functions plus $g$ times sums of exponential functions converging to $f$ in $||.||_\infty$.
\end{lemma}
{\bf proof:} We use that there is a unique split $f=f_a+f_o$ with $f_a\in AP(\mathbb R)$ and $f_o\in C_o(\mathbb R)$. Using the density of the exponential functions $AP(\mathbb R)$ we find a Cauchy sequence of finite sums of exponential functions converging to $f_a$. \\
For $f_o$ we use that for every $\epsilon>0$ there exists $a,b\in \mathbb R$ and $\tilde{f} \in C_c(\mathbb R)$ with $\text{supp}(\tilde{f})\subset [a,b]$ s.t. $||f-\tilde{f}||_\infty \le \epsilon$. Using $\tilde{f}=\chi_{a,b} \tilde{f}_{a,b}$, we find that for every $\epsilon >0$ there exists $a,b$ and and $\tilde f_{a,b} \in AP(\mathbb R)$ s.t. $|| f_o- \chi_{a,b} \tilde{f}_{a,b}|| \le \epsilon$. Moreover, using $g(x)>0$ and $g(x)\le\frac{g_o}{x^2}$ for $x>x_o$, we find for $\epsilon>0$ there exist $A,B\in \mathbb R$ s.t. \linebreak  $||g\,G^{A,B}_{a,b}- \chi_{a,b} ||_\infty\le \epsilon$. Hence from
$$
  ||f_o-g 
G^{A,B}_{a,b}\tilde f_{a,b}||_\infty 
= || f_o-(g\, G^{A,B}_{a,b}-\chi_{a,b}+\chi_{a,b})\tilde{f}_{a,b} ||\le \epsilon\left(1+||\tilde f_{a,b}||_\infty\right)
$$
and $||\tilde f_{ab}||_\infty\le ||f_o||_\infty +\epsilon$ and the density of the exponential functions in $AP(\mathbb R)$ we find that for every $\epsilon^\prime>0$ there exists a sum of exponentials $h$ yielding $||f_o-g h||_\infty\le \epsilon^\prime$.
$\blacksquare$
\paragraph{An Explicit Embedding.} Now we use this result in order to construct an explicit embedding of $AAP(\mb{R})$ into $Cyl(\ol{\MC{A}})$, the configuration space of full LQG. For this we use the spiral arc solution of section \ref{Non-Embedability of Configuration Spaces} and the defining representation of $SU(2)$. 
First we construct a function $g(c)>0$ with $g(c)\le\frac{g(c_0)}{c^2}$ $\forall c>c_0$. 
Recall the solution (\ref{final ODE solution spiral Case 2})
\[
     \mathfrak{a}(\kappa,t)
   =\mb{e}^{\frac{\mb{i}}{2}\lambda t}
   \left\{\cos(\Delta t)+\frac{\mb{i}}{2\Delta}(2\nu c -\lambda)\sin(\Delta t)\right\}
\]
where  $\Delta=\big[\frac{\lambda^2}{4}+c(c-\nu\lambda)\big]^{1/2}$ and $ \lambda,\nu\in\mb{R}, \mu\in\mb{C}$, $\nu^2+|\mu|^2=1$. Now denote the explicit dependence on the parameters $\mu,\nu,\lambda,t$ as subscripts $\mathfrak{a}_{\nu,\lambda,t}$ (because $\mathfrak a$ is independent of $\mu$) and compute
\begin{equation}\label{equ:Bdefi}
 \begin{array}{rcl}
   B_{\lambda,t}(c)&=&\left|\mathfrak{a}_{0,0,t}(0)-e^{\frac i 2 \lambda t}\mathfrak{a}_{0,\lambda,t}(c)\right|^2\\
     &=& \frac{\lambda ^2}{4 c^2+\lambda ^2}\sin ^2\left(\sqrt{c^2+\frac{\lambda ^2}{4}}t\right)+\left(\cos (c t)-\cos \left(\sqrt{c^2+\frac{\lambda
   ^2}{4}}t\right)\right)^2
 \end{array}
\end{equation}
as well as
\begin{equation}\label{equ:Adefi}
 A_{t}(c)=\mathfrak{a}_{1,0,t}(c)=e^{i c t}.
\end{equation}
We verify for $t>0$ and $\lambda>0$ that $B_{\lambda,t}(c)=\mathcal O(c^{-2})$ for $c\to \infty$, by inspecting both summands separately: First, $\frac{\lambda ^2}{4 c^2+\lambda ^2}\sin ^2\left(\sqrt{c^2+\frac{\lambda ^2}{4}}t\right)=\mathcal O(c^{-2})$, because the modulus of the sine is bounded by 1 and the quotient is $\mathcal O(c^{-2})$ as $c\to \infty$. Second, $\left(\cos (c t)-\cos \left(\sqrt{c^2+\frac{\lambda^2}{4}}t\right)\right)^2=\mathcal O(c^{-2})$ because expanding the square root inside the argument of the second cosine around 1 cancels to first order the first cosine and the remainder is of $\mathcal O(c^{-1})$ which is squared to yield $\mathcal O(c^{-2})$. Moreover, for $\lambda=2$ and $t=1$ we verify $B_{2,1}(c)>0$ as follows: Both summands are squares and thus positive semidefinite. The first summand vanishes for $c=\frac 1 2 k \pi-\frac 1{2 k \pi}$ for integer $k$, while inserting these values for $c$ into the argument of the sine is incompatible with $\frac n 2 \pi$ for integer $n$.  
Hence $B_{2,1}(c)$ can be used for the construction in the proof of lemma \ref{lem:AAPapproximation}.

To construct cylindrical functions in LQG whose restriction to homogeneous isotropic connections reduces to the functions (\ref{equ:Adefi}) and (\ref{equ:Bdefi}) of the homogeneous isotropic connection parameter $c$, we realize $\mathfrak{a}$ as the holonomy matrix element $h_{11}$ and $\bar{\mathfrak a}$ as the holonomy matrix element $h_{22}$ in the fundamental representation. Moreover, we fix three general points $\xi,\upsilon,\zeta$, which we denote as superscripts e.g. $h^\xi_{ab}$, and specify spiral parameters $\nu,\lambda,t$, which we denote as arguments $h^\xi_{ab}(\nu,\lambda,t)$ and define the embedding map $r$ as the linear extension of 
\begin{equation}\label{equ:r1}
 r(1)=1
\end{equation}
\begin{equation}\label{equ:r2}
 r(e^{\i c t})=h^\xi_{11}(1,0,t)
\end{equation}
\begin{equation}
 r(B_{2,1}(c)e^{\i c t})=h^\upsilon_{11}(1,0,t)\left(h^\zeta_{11}(0,0,1)-e^{-\i}h^\upsilon_{11}(0,2,1)\right)\left(h^\zeta_{22}(0,0,1)-e^\i h^\upsilon_{22}(0,2,1)\right),
\end{equation}
where $t\ne 0$ in (\ref{equ:r2}) and the $1$ in (\ref{equ:r1}) is understood as the unit function.

\paragraph{An Explicit Measure for the Symmetry Reduced Sector.}
Let us now consider the image of $r$ within the Loop Quantum Gravity inner product:
Observing that $r(e^{ict})$ is (upon normalization) a gauge-variant spin network function on a different graph for different $t$, we conclude 
\begin{equation}
  \frac 1{|r(e^{\i ct})|_{LQG}|r(e^{\i cs})|_{LQG}} \langle r(e^{\i ct}), r(e^{\i cs}) \rangle_{LQG}=\delta^{Kr.}_{t,s},
\end{equation}
where we used $|T|_{LQG}=\sqrt{\langle T,T \rangle_{LQG}}$ and the Kronecker delta $\delta^{Kr.}_{s,t}$. Moreover, observing that $r(B_{2,1}(c)\,e^{\i t c})$ is a linear combination of gauge variant spin network functions on a graph always different from $r(e^{\i s c})$ and for $s\ne t$ on a different graph than $r(B_{2,1}(c)\,e^{\i s c}))$, we find
\begin{equation}
  \frac 1{|r(B_{2,1}(c)\,e^{\i t c})|_{LQG}|r(e^{\i s c})|_{LQG}} \langle r(B_{2,1}(c)\,e^{\i t c}), r(e^{\i s c}) \rangle_{LQG}=0
\end{equation}

\begin{equation}
 \frac 1{|r(B_{2,1}(c)\,e^{\i t c})|_{LQG}|r(B_{2,1}(c)\,e^{\i s c})|_{LQG}} \langle r(B_{2,1}(c)\,e^{\i t c}), r(B_{2,1}(c)\,e^{\i s c}) \rangle_{LQG}=\delta^{Kr.}_{t,s}.
\end{equation}
We have thus established that orthogonality of our generating set and can return to the question of completion: For this purpose let us introduce the induced norm $f$ in the span of the generating set: 
\begin{equation}
 ||f||_{ind}:=|r(f)|_{LQG},
\end{equation}
which we can freely use to construct a completion of the span of the generating set. However, we have to caution that we where not able to prove that the completions in $||.||_\infty$ is contained in the completion using $||.||_{ind}$. Hence we can not exclude the possibility that the embedding map $r$ needs to be amended.

\subsection{\label{Asymptotic Analysis and the BKL Picture}Asymptotic Analysis and the BKL Picture}

Let us briefly discuss some physical consequences of the construction described in the previous two subsections:

The BKL-conjecture \cite{Belinsky:1970ew,Khalatnikov:1969eg,Belinsky:1982pk} suggests that the generic evolution of general relativity near a singularity is such that spatial derivatives become negligible compared to time derivatives, i.e. each point in space evolves separately as a homogeneous cosmology, so the understanding of cosmological models seems to hint towards an understanding of the evolution near a generic singularity. Since large time derivatives of the metric imply large values of the Ashtekar connection in a fixed trivialization and a fixed chart, a first approach for the investigation of the behavior of loop quantum gravity observables near a generic singularity is to investigate the asymptotic dependence of observables and in particular spin network functions on homogeneous connections. Thus our result, stating that the dependence of a spin network function $T(A)$ on the homogeneous scaling parameter $c$ of a connection $A=c A_*$ becomes asymptotically almost periodic as $c\to\infty$, suggests that the generic behavior of spin network functions near a singularity is well approximated by an almost periodic dependence on the scaling parameter $c$. This scenario has been investigated in the context of LQC \cite{Ashtekar2006,Ashtekar2008,Ashtekar2009,Ashtekar2010}.

However in light of the nonembeddability of standard LQC into LQG one expects modifications to the standard LQC dynamics, since it is expected that quantum corrections will become important not at infinite momentum scale, but rather at momentum scale of order $1$ in Planck units, one expects from dimensional analysis
\begin{equation}
 T(c A_*) = \psi(c) (1+\mathcal O(1/c)),
\end{equation}
where $\psi$ is almost periodic, that the $\mathcal O(1/c)$ corrections are not necessarily negligible when approaching the Planck scale and become negligible only when going far beyond Planck scale. It is thus possible that the use of definition \ref{defi:quantum-symmetry} could lead to observable deviations from results derived using definition \ref{defi:classical-symmetry}. However the existence of quantum connections of the type defined in equation \ref{equ:funky-symmetry} suggests to use definition \ref{defi:quantum-symmetry}.

\section{\label{Conclusions}Conclusions and Outlook}

Loop quantum cosmology (LQC) is used as a tool for the study of conceptual issues in full loop quantum gravity (LQG), because it is a simple toy model that exhibits many features of the full theory. Moreover, since on the one hand it is expected that quantum effects of gravity are most significant in the vicinity of general relativity (GR) singularities and on the other hand since the BKL conjecture \cite{Belinsky:1970ew,Khalatnikov:1969eg,Belinsky:1982pk} states that the dynamics of GR in the vicinity of a generic singularity is well approximated by decoupled homogeneous cosmologies, one expects that LQC provides insight into the behavior of LQG where its effects are expected to be most important. However, to be able to draw such conclusions, one needs to understand the relation between full LQG and the symmetry reduced model. The nature of this relation is not as obvious as one might think, because LQC is obtained as a \lq\lq loop quantization\rq\rq of a classical symmetry reduction of GR and not as a symmetry reduction of LQG. More disturbingly, one finds that the structures of the two theories do not match up: In particular one can show that the embedding of the configuration space of LQC into the configuration space of LQG is not continuous \cite{Brunnemann:2007du}. Since LQG is a rather restrictive field theoretic framework that does not admit arbitrary changes, we are lead to the investigation of symmetry reduced models of LQG in this paper.

It turns out that the problem of non-embeddability of standard LQC into full LQG stems from the restriction to piecewise linear curves in the \lq\lq loop quantization\rq\rq of homogeneous cosmology, which in turn implies that the holonomy matrix elements, which are the building blocks of the configuration operators in the \lq\lq loop quantization\rq\rq program are linear combinations of exponential functions of the symmetric connection parameter. This motivated our first Ansatz to start with standard LQC, but to allow for small perturbations of the linear curves used in the construction. To do this we formally solved the holonomy differential equation as a power series in the perturbation parameter, such that the first term in the series is precisely the unperturbed linear curve. This provides an intuitive understanding of the relation of standard LQC and its embeddable extension. Moreover, it turns out that the formal solution can not converge for large values of the symmetric connection parameter unless the perturbation parameter scales with the inverse of the symmetric connection parameter. 

This inverse scaling motivated our second Ansatz, the asymptotic expansion of the solution to the holonomy differential equation as an inverse power series in the symmetric connection parameter. The resulting Liouville Green series is in general divergent, but one can show that the truncated Liouville Green series captures the asymptotic behavior correctly. It follows that the solutions to the holonomy differential equation can be characterized as a sum of a continuous periodic function and a continuous function that vanishes at infinity, implying that the algebra generated by polynomials thereof can be characterized as continuous asymptotically almost periodic function. It turns out that this result generalizes from symmetric connections to {\it {arbitrary}} one-parameter families of connections of form $A=c A_*$, where spin network functions can be shown to be asymptotically almost periodic functions of the parameter $c$.

In the following we provided a definition of a symmetric quantum connection that does {\it not} tacitly assume classical smoothness properties of a quantum connection. The asymptotic almost periodic dependence of configuration operators on the symmetric connection parameter shows that one has to include at least one nonlinear curve in the construction of an embeddable symmetry reduced model of LQG and also allowed us to give an explicit construction of an embeddable symmetry reduced model of full LQG, that contains standard LQC and a \lq\lq minimal\rq\rq extension. This extension consists, in accordance with \cite{Fleischhack2010}, of configuration observables that vanish both, for large and small values of the symmetric connection parameter.

It is likely that the extension does not change the generic physical behavior in the limit of infinite symmetric connection parameter that was established in LQC, because it consists of functions that vanish for large connection parameters. Thus, elements of the extension are not expected to affect the behavior of the model in the limit of infinite connection parameter. Specific physical predictions are however likely to change, because quantum gravity effects are expected to play an important role not at infinite connection parameter but at values of order one (or even less) in Planck units, where the configuration operators introduced through the extension do not vanish. In particular, if we are using the usual configuration representation, there are wave functions in the extension that vanish for large connection parameter, or physically speaking near the cosmological singularity, which are to be expected to yield very different physical behavior than almost periodic wave functions.

\subsection{\label{Outlook}Outlook}

The results presented in this paper shed new light on the conceptual question what symmetry means at the level of the kinematical LQG-Hilbert space $\Ho$ and how symmetric states can be constructed thereon. The novel answer we get from our analysis is that every specifically chosen symmetric connection $cA_*$ induces a decomposition of $\Ho$ into equivalence classes of cylindrical functions, where two cylindrical functions are called equivalent, if their restriction to $cA_*$ coincides. Hence we can take the opposite point of view and {\it define} a particular symmetry by an according decomposition of $\Ho$ into equivalence classes of cylindrical functions.
  
Starting from this observation there are several future directions for investigation. 

An important task is to examine the mathematical properties of the symmetry-equivalence classes in more detail. The ultimate goal here is to work out the general construction of a symmetric Hilbert space $\MC{H}_{\text{sym}}$ by taking the quotient of $\Ho$ and the equivalence relation\footnote{This generalizes to very general symmetries (e.g. discrete symmetries) and the questions raised here are applicable in these cases as well.}. It will be important to generalize the construction of section \ref{ExplQuantSymm} and to see how a measure on $\MC{H}_{\text{sym}}$ can be constructed and to understand, whether the symmetry equivalence classes preserve the inductive structure of $\Ho$.
 
Secondly, given the explicit embedding of the extended LQC-configuration space into $\Ho$ as described in section \ref{ExplQuantSymm} we have already proved that the $AAP(\mb{R})$-functions form an algebra. However one needs to work out the action of flux operators of full LQG on a particular embedding of $AAP(\mb{R})$ in $\ol{\MC{A}}$. Does this action preserve the symmetric sector and moreover the symmetry equivalence classes in $\Ho$? If it does, the action of geometric operators and ultimately the constraint operators on the symmetry-equivalence classes has to be analyzed in detail. These questions can be investigated very naturally using the results of this paper, because they suggest to define symmetric operators as operators on the LQG Hilbert space as follows: Recall that we have an equivalence relation $\sim$ of cylindrical functions through the pull-back under $i:c\mapsto c A_*$, such that any cylindrical function can be written as the sum of a fixed representative $f_\sim$ in the equivalence class plus a function $f^o$ whose pull-back under $i$ vanishes. We then call an operator $O$ symmetric if it satisfies
\begin{equation}
 \langle Cyl_1, O Cyl_2 \rangle = \langle (Cyl_1+f_1^o), O (Cyl_2+f^o_2) \rangle,
\end{equation}
for all cylindrical functions $Cyl_i$ and cylindrical functions $f^o_i$ whose pull-back under $i$ vanishes. Notice that this relation is by construction linear, meaning that if $O_1,O_2$ are symmetric, then $O_1+\alpha O_2$ for $\alpha\in \mathbb C$ is symmetric as is $O_1 O_2$.

A third very important goal is to analyze the effect the extension of the configuration space of LQC from $\ol{\mb{R}}_{\text{Bohr}}$ to the spectrum  $\Delta(AAP(\mb{R}))$ has. Certainly the first goal is to construct a measure on the extended Hilbert space. Beside starting from the extended configuration space itself there is the possibility to start from our explicit embedding and its usage of the kinematical inner product on $\Ho$ using the Ashtekar-Lewandowski measure. As described above  one might be able to define a measure on $\MC{H}_{\text{sym}}$ and to use it for the extended version of LQC. Then one is able to work out the consequences for the physical predictions of $LQC$.

Lastly, there are some purely mathematical questions: From the viewpoint of differential geometry it will be interesting to see, how our procedure (described in section \ref{Generalization to the Full Configuration Space})  for computing the holonomy of a connection in the asymptotic regime of a large  re-scaling can be generalized to connections on arbitrary bundles. Moreover it is necessary to work out its relation to the curvature of the connection as discussed in section \ref{Interpretation of the Symmetric Connection} in more detail. From a complex analysis viewpoint one is lead to ask when solutions to the holonomy equations are analytic in $c$, this could be investigated e.g. through extending the constructions of section \ref{ExplQuantSymm} to complex $c$ and  applying results of \cite{Dunster1993}. From a functional analytic viewpoint one needs to address the convergence properties of the embedding map, e.g. through proving bounds that imply convergence.

\addcontentsline{toc}{section}{\numberline{}Acknowledgments}
\section*{Acknowledgments}
The work of JB has been supported by the Emmy-Noether-Programm (grant FL 622/1-1) of the Deutsche Forschungsgemeinschaft. TK was supported by the Perimeter Institute for Theoretical Physics. Research at the Perimeter Institute is supported in part by the Government of Canada through NSERC and by the Province of Ontario through MEDT. JB thanks  F. Schulte-Hengesbach and the Department of Mathematics at Hamburg university, where parts of this work have been carried out, for hospitality. JB thanks the Perimeter Institute for hospitality. We thank Christian Fleischhack and T.~M.~Dunster for discussions. 

\pagebreak

\begin{appendix}
\section{\label{Appendix epsilon schranke} Boundedness of Formal Solution}

By construction the series (\ref{final A definition})  provides a formal solution to the ODE (\ref{LG-Vorbereitung 2}). However it is not clear whether this series converges.
we thus truncate it and investigate the rest term 
\barr{lcl}
  \epsilon^{(\l)}_{n>0}(\kappa,t)
  &:=&\displaystyle
  d^{(\l)}(\kappa,t)-\mb{e}^{\l\kappa t+\frac{\l}{2}\int_0^t \alpha(s)\,ds}\sum_{k=0}^{n-1}\frac{A^{(\l)}_k(t)}{\kappa^k}
\earr
separately. 
Insertion into the holonomy ODE yields
\begin{equation}\label{equ:epsilon-ODE}
  \begin{array}{rcl}
    \ddot \epsilon^{(\l)}_n 
    &=& (\kappa^2+\kappa \alpha+\beta )\epsilon_n^{(\l)}
    +\frac{\mb{e}^{\l \kappa t}}{\kappa^{n-1}}\Xl
    \left(\beta  A_{n-1}^{(\l)}-\left(\frac{\ddot X^{(\l)}}{\Xl}A^{(\l)}_{n-1}
                               +2 \frac{\dot X^{(\l)}}{\Xl}\dot A^{(\l)}_{n-1}
                               +\ddot A^{(\l)}_{n-1} \right)\right)\\
      &=& (\kappa^2+\kappa \alpha+\beta )\epsilon_n^{(\l)}+\Delta^{(\l)}_n.
  \end{array}
\end{equation}
where $\Delta^{(\l)}_n(\kappa,t)= \mb{e}^{\l \kappa t}\Xl(t) \frac{ 2\l \dot A^{(\l)}_{n}(t)}{\kappa^{n-1}} $. To investigate the asymptotic behavior of $\epsilon^{(\l)}_n$, we construct a fundamental system with Wronskian $2$ that solves equation (\ref{equ:epsilon-ODE}) to all positive powers in $\kappa$
\begin{equation}
  Z_{(\sigma)}(\kappa,t)=\frac{\mb{e}^{\sigma\left(\kappa t+\frac 1 2 \int_0^t  \,\alpha(\tau)\,d\tau\right)}}{\sqrt{\kappa+\frac 1 2 \alpha(t)}},
\end{equation}
with $\sigma=\pm 1$, which satisfies 
\begin{equation}
  \ddot Z_{(\sigma)}=\left(\kappa^2+\kappa\alpha+\gamma\right)Z_{(\sigma)},
\end{equation}
where $\gamma(\kappa,t)=\frac{\alpha^2(t)}{4}
                       +\frac{3\dot \alpha^2(t)}{16(\kappa+ \frac 1 2 \alpha(t))^2}
                       -\frac{\ddot \alpha(t)}{4(\kappa+\frac 1 2 \alpha(t))}$. Variation of constants of
\begin{equation}
  \epsilon^{(\l)}_n=\mu^{(\l)}_n Z_{(+)}+\nu^{(\l)}_n Z_{(-)}
\end{equation}
yields
\begin{equation}
  \begin{array}{rcl}
    \dot \mu^{(\l)}_n Z_{(+)}+\dot \nu^{(\l)}_n Z_{(-)} &=&0\\
    \dot \mu^{(\l)}_n \dot Z_{(+)}+\dot \nu^{(\l)}_n \dot Z_{(-)}&=&(\beta-\gamma)\epsilon^{(\l)}_n+\Delta^{(\l)}_n,
  \end{array}
\end{equation}
yielding for $\epsilon^{(\l)}_n(\kappa,t)$
\begin{equation}
    \epsilon^{(\l)}_n(\kappa,t)=\epsilon^{(\l)}_n(\kappa,0)+ \int_0^t ~ K(t,\tau,\kappa)\left(\chi(\kappa,\tau)\epsilon^{(\l)}_n(\kappa,\tau)+\Delta^{(\l)}_n(\tau)\right)~d \tau~~,
\end{equation}
where 
$K(t,\tau,\kappa)=\frac{1}{2}
\left(Z_{(+)}(\kappa,t)Z_{(-)}(\kappa,\tau)-Z_{(-)}(\kappa,t)Z_{(+)}(\kappa,\tau)\right)$ and $\chi(\kappa,\tau)=\beta(\tau)-\gamma(\kappa,\tau)$.
Imposing the initial condition $\epsilon^{(\l)}_n(\kappa,0)=0,\,\dot\epsilon^{(\l)}_n(\kappa,0)=0$ on the integral equation, we can construct a series solution given by $\epsilon^{(\l)}_n=\sum_{m=0}^\infty \left(\epsilon^{(\l)}_{n,m+1}-\epsilon^{(\l)}_{n,m}\right)$, where the summands are determined by
\begin{equation}
  \begin{array}{rcl}
    \epsilon^{(\l)}_{n,0}(\kappa,t)
    &=&0
    \\
    \epsilon^{(\l)}_{n,1}(\kappa,t)
    &=&\displaystyle
    \int_0^t d\tau\, K(t,\tau,\kappa)\,\Delta^{(\l)}_n(\kappa,\tau)
    \\
    \epsilon^{(\l)}_{n,m+1}(\kappa,t)-\epsilon^{(\l)}_{n,m}(\kappa,t)
    &=&\displaystyle
    \int_0^t d\tau\, K(t,\tau,\kappa)\,\chi(\kappa,\tau)\left(\epsilon^{(\l)}_{n,m}(\kappa,\tau)-\epsilon^{(\l)}_{n,m-1}(\kappa,\tau)\right).
  \end{array}
\end{equation}
To find a bound on this series, we impose
\begin{equation}
  \begin{array}{rcl}
    |K(t,\tau,\kappa)|&\le&p(\kappa,t)\,q(\tau)\\
    \MC{K}&:=&\sup_{0\le\tau\le t}\{|q(\tau)|\}\\
    \MC{K}_o(\kappa)&:=&\sup_{0\le\tau\le t}\{|p(\kappa,\tau)\,q(\tau)|\}\\
    \phi^{(\l)}_n(\kappa,t)&:=&\displaystyle\int_0^t  |\Delta^{(\l)}_n(\kappa,\tau)|\,d\tau\\
    \psi(\kappa,t)&:=&\displaystyle\int_0^t |\chi(\kappa,\tau)|\,d\tau,
  \end{array}
\end{equation}
where $p,q\ge 0$, such that we can show
\begin{equation}
  \frac{|\epsilon^{(\l)}_{n,m+1}(\kappa,t)-\epsilon^{(\l)}_{n,m}(\kappa,t)|}{p(\kappa,t)}
  \le \MC{K} \,\phi^{(\l)}_n(\kappa,t) \frac{\MC{K}_o^m\psi^m(\kappa,t)}{m!}
\end{equation}
by complete induction. Induction beginning:
\begin{equation}
  \begin{array}{rcl}
    \big|\epsilon^{(\l)}_{n,1}(\kappa,t)-\epsilon_{n,0}^{(\l)}(\kappa,t)\big|
    &=&\displaystyle
    \big|\epsilon^{(\l)}_{n,1}(t)|\le\int_0^t d\tau\big|K(t,\tau,\kappa)\big|\big|\Delta^{(\l)}_n(\kappa,\tau)\big|
    \\
    &\le&\displaystyle
    p(\kappa,t)\int_0^td\tau\, q(\tau)\big|\Delta^{(\l)}_n(\kappa,\tau)\big|
    \\
    &\le&\displaystyle
    p(\kappa,t)\,\MC{K}\, \int_0^t d\tau\,\big|\Delta^{(\l)}_n(\kappa,\tau)\big|\,
    \\
    &\le&\displaystyle
    p(\kappa,t)\, \MC{K}\, \phi^{(\l)}_n(\kappa,t).
  \end{array}
\end{equation}
Induction step:
\begin{equation}
  \begin{array}{rcl}\displaystyle
    \frac{|\epsilon^{(\l)}_{n,m+1}(\kappa,t)-\epsilon^{(\l)}_{n,m}(\kappa,t)|}{p(\kappa,t)}
    &=&\displaystyle
    \frac{1}{p(\kappa,t)}\left|\int_0^t d\tau K(t,\tau,\kappa)\chi(\kappa,\tau)\left(\epsilon^{(\l)}_{n,m}(\kappa,\tau)-\epsilon^{(\l)}_{n,m-1}(\kappa,\tau)\right)\right|
    \\
    &\le&\displaystyle
    \frac{1}{p(\kappa,t)}\int_0^t d\tau\big|K(t,\tau,\kappa)\big|\big|\chi(\kappa,\tau)\big| \big|\epsilon^{(\l)}_{n,m}(\kappa,\tau)-\epsilon^{(\l)}_{n,m-1}(\kappa,\tau)\big|
    \\
    &\le&\displaystyle
    \int_0^t d\tau\, q(\tau)\big|\chi(\kappa,\tau)\big|\big|\epsilon^{(\l)}_{n,m}(\kappa,\tau)-\epsilon^{(\l)}_{n,m-1}(\kappa,\tau)\big|
    \\
    &\le&\displaystyle
    \int_0^t d\tau\, q(\tau)\big|\chi(\kappa,\tau)\big|\,\MC{K}\, \phi^{(\l)}_n(\kappa,\tau)\frac{\MC{K}_o^{m-1}\psi^{m-1}(\kappa,\tau)}{(m-1)!}p(\kappa,\tau)
    \\
    &\le&\displaystyle
    \MC{K}\,\MC{K}_o^m(\kappa) \int_0^t d\tau \big|\chi(\kappa,\tau)\big|\,\phi^{(\l)}_n(\kappa,\tau)\frac{\psi^{m-1}(\kappa,\tau)}{(m-1)!}
    \\
    &\le&\displaystyle
    \MC{K}\, \MC{K}_o^m(\kappa)\,\phi^{(\l)}_n(\kappa,t)\int_0^t d\tau\big|\chi(\kappa,\tau)\big|\frac{\psi^{m-1}(\kappa,\tau)}{(m-1)!}
    \\
    &=&\displaystyle
    \MC{K}\, \MC{K}_o^m(\kappa)\,\phi^{(\l)}_n(\kappa,t)\frac{\psi^m(\kappa,t)}{m!}.
  \end{array}
\end{equation}
where we used the fact that by construction $\phi^{(\l)}_n(\kappa,t)$ and $\psi(\kappa,t)$ are non decreasing in the interval $[0,t]$. The series is then bounded by
\begin{equation}
  \begin{array}{rcl}
    \big|\epsilon^{(\l)}_n(\kappa,t)\big|
    &=&\displaystyle
      \left|\sum_{m=0}^\infty\left(\epsilon^{(\l)}_{n,m+1}(\kappa,t)-\epsilon^{(\l)}_{n,m}(\kappa,t)\right)\right|
      \\
      &\le& \displaystyle
      p(\kappa,t)\ \MC{K}\, \phi^{(\l)}_n(\kappa,t)\sum_{m=0}^\infty \frac{\big(\MC{K}_o(\kappa)\,\psi(\kappa,t)\big)^m}{m!}
      \\
      &=&\displaystyle
      p(\kappa,t)\,\MC{K} ~\phi^{(\l)}_n(\kappa,t)~ \mb{e}^{\MC{K}_o(\kappa)~ \psi(\kappa,t)}.
  \end{array}
\end{equation}
Using $\lambda(\kappa,t):=\inf_{0\le\tau\le t}\{|\kappa+\frac 1 2 \alpha(\tau)|\}$, which is $\mathcal O(\kappa)$ as $|\kappa|\to\infty$, we can bound $K(t,\tau,\kappa)$ by
\begin{equation}
  \begin{array}{rcl}
    \big|K(t,\tau,\kappa)\big|
    &=&\displaystyle
    \frac 1 2
    \Big|\left(Z_{(+)}(\kappa,t)Z_{(-)}(\kappa,\tau)-Z_{(-)}(\kappa,t)Z_{(+)}(\kappa,\tau)\right)\Big|
    \\[2mm]
    &=&\displaystyle
    \frac{1}{2}
    \Big|\frac{
                \mb{e}^{\kappa(t-\tau)+\frac{1}{2}\int_{\tau}^t\alpha(s)ds}
               -\mb{e}^{-\kappa(t-\tau)-\frac{1}{2}\int_{\tau}^t\alpha(s)ds}
              }
              {\sqrt{\big(\kappa+\frac{1}{2}\alpha(t)\big)
                     \big(\kappa+\frac{1}{2}\alpha(\tau)\big)}
              }
    \Big|
    \\
    &\le&\displaystyle
    \frac{1}{2\,\lambda(\kappa,t)}
    \Big(      \big| \mb{e}^{\kappa(t-\tau)+\frac{1}{2}\int_{\tau}^t\alpha(s)ds}\big| 
              +\big| \mb{e}^{-\kappa(t-\tau)-\frac{1}{2}\int_{\tau}^t\alpha(s)ds}\big| 
    \Big)
    \\[2mm]
    &=&\displaystyle
    \frac{1}{2\,\lambda(\kappa,t)}
    \Big(      \big| \mb{e}^{\frac{1}{2}\int_{\tau}^t\alpha(s)ds}\big| 
              +\big| \mb{e}^{-\frac{1}{2}\int_{\tau}^t\alpha(s)ds}\big| 
    \Big)
    \\
    &\le&\displaystyle
    \frac{1}{2\,\lambda(\kappa,t)}
    \Big(      \big| \mb{e}^{\frac{1}{2}\int_{\tau}^t|\alpha(s)|ds}\big| 
              +\big| \mb{e}^{\frac{1}{2}\int_{\tau}^t|\alpha(s)|ds}\big| 
    \Big)
    \\
    &=&\displaystyle
    \frac{\mb{e}^{\frac{1}{2}\int_{\tau}^t|\alpha(s)|ds}}{\,\lambda(\kappa,t)}
    \\
    &=&\displaystyle
    \frac{\mb{e}^{\frac{1}{2}\int_{0}^t|\alpha(s)|ds}}{\lambda(\kappa,t)}
    \cdot \mb{e}^{-\frac{1}{2} \int_{0}^{\tau} |\alpha(s)|ds}
   .
  \end{array}
\end{equation}
Here we have used that $\kappa=\mb{i}c$ ($c\in\mb{R}$) is purely imaginary. We can thus choose
\begin{equation}
  \begin{array}{rcl}
    p(\kappa,t)&=&\frac{\exp\left(\frac 1 2\int_0^t |\alpha(s)|ds\right)}{\lambda(\kappa,t)}\\
    q(\tau)&=&\exp\left(-\frac 1 2 \int_0^\tau |\alpha(s)|\,ds\right),
  \end{array}
\end{equation}
which allows us to set
\begin{equation}
  \begin{array}{rcl}
    \MC{K}&=&\sup_{0\le\tau\le t}\left\{|q(s)|\right\}
           = \sup_{0\le\tau\le t}\left\{\left|\mb{e}^{-\frac 1 2\int_0^\tau |\alpha(s)|ds}\right|\right\}\le \exp\left(\frac 1 2 \int_0^t |\alpha(s)|\,ds\right)\\
    \MC{K}_o(\kappa)&=&\sup_{0\le \tau\le t}\{|p(\kappa,\tau)\,q(\tau)|\}=
    \frac{\exp\left(\frac 1 2\int_0^t |\alpha(s)|ds\right)}{\lambda(\kappa,t)}.
  \end{array}
\end{equation}
We thus established the bound
\be\label{bound on truncation error}
  \big|\epsilon^{(\l)}_n(\kappa,t)\big|
  \le 
  \frac{\exp\left(\int_0 ^t |\alpha(s)|\,ds\right)}{\lambda(\kappa,t)}
  \left(\int_0^t \big|\Delta^{(\l)}_n(\kappa,\tau)\big|\,d\tau\right)
           \exp\left(\exp\left(\frac 1 2\int_0^t |\alpha(s)|ds\right) \frac{\int_0^t |\chi(s)|\,ds}{\lambda(\kappa,t)}\right)\,,
\ee
that ensures that $\epsilon^{(\l)}_n(\kappa,t)\to 0$ as $|\kappa|\to \infty$ if $\alpha(t),\Delta^{(\l)}_n(\kappa,t),\chi(\kappa,t)$ are bounded.

\subsection{\label{Estimate of the Error Term:Dunster}Estimate of the Error Term}

Let us assume that $\psi,\alpha$ are analytic in a domain $D(d)$ that contains all complex numbers $z$ for which $\min_{x\in I}\{|z-x|\le d\}$, where $I$ is the integration path for the Liouville-Green expansion and furthermore assume that there are constants $k_1,k_2,A$ s.t. for all $x \in D(d)$ we have the bounds
\begin{equation}
 |\psi(x)|\le k_1,\,\int_o^t ds |\psi(x)|\le k_2\,|\alpha(x)|\le A,
\end{equation}
so we can define $k=\max\{k_1,\frac{k_2}{d},\frac{A}{d}\}$. The bounds yield $|\psi^{(m)}(t)|\le k_1 \frac{m!}{d^m}$ and $|\alpha^{(m)}(x)|\le A \frac{m!}{d^m}$ through Cauchy's integral formula. 
\begin{lemma}\label{lemma:coefficient-bounds}
 Given the above bounds, the Liouville-Green coefficients satisfy for $t$ within the integration path and $n\ge 2$: $|d_n^{(m)}(t)|\le C_n \frac{k}{2^n}\frac{(m+n-2)!}{d^{m+n-2}}$ with $C_n=(1+2 k d^2)\frac{(1+2 k d^2)_{n-2}}{(n-2)!}$. 
\end{lemma}
{\bf proof:} by complete induction following \cite{Dunster1993}. preparation:
\begin{equation}
 \begin{array}{rcl}
  |d_1(t)|&=&\displaystyle\frac 1 2 \Big|\int_o^t du \psi(u\Big)|\le \frac 1 2 k_2\\
  |d_1^{(m)}|&=&\displaystyle\frac 1 2  \Big|\psi^{(m-1)}(t) \Big|\le \frac 1 2 k_1 \frac{(m-1)!}{d^{m-1}}.
 \end{array}
\end{equation}
Induction beginning:
\begin{equation}
 \begin{array}{rcl}
   |d_2(t)|&\le&\displaystyle\frac 1 2 \int_o^t du |d_1(u)||\psi(u)|+\frac 1 4 |\psi(t)|+\frac 1 2 \int_o^tdu |\alpha(u)||d_1^{(1)}(u)|\\
           &\le&\displaystyle\frac 1 4\left(k_2^2+ k_1+A k_2\right)\le\frac k 4\left(1+2k d^2\right).
 \end{array}
\end{equation}
Furthermore by denoting the digamma function by $\psi$ and using $\sum_{k=1}^{m-1}\frac{1}{k(m-k)}=2\frac{\gamma+\psi(m)}{m}$ as well as $\frac 1 m+2\frac{\gamma+\psi(m)}m\le 1$ for all natural numbers $m$ we find
\begin{equation}
 \begin{array}{rcl}
   |d^{(m)}_2(t)|&\le&\displaystyle\frac 1 2 \Big|\frac{d^m}{dt^m}(d_1(t))^2\Big|
   +\frac 1 4 \Big|\psi^{(m)}(t)\Big|+\frac 1 2 \Big|\frac{d^{m-1}}{dt^{m-1}}(\alpha(t) d_1(t))\Big|\\
     &\le&\displaystyle \frac 1 4 k \frac{m!}{d^m}+\frac 1 2\left(2 |d_1^{(m)}(t)||d_1(t)|+\sum_{k=1}^{m-1}\left(m\atop k\right)|d_1^{(m-k)}(t)||d_1^{(k)}(t)|\right)\\
     &&\displaystyle+\frac 1 2 \sum_{l=0}^{m-1}\left(m-1\atop l\right)A\frac{(m-l-1)!}{d^{m-l-1}}\frac{k_1}2 \frac{l!}{d^l}\\
     &\le&\displaystyle\frac 1 4 k_3 \frac{m!}{d^{m}}\left(1+k \frac{d^2}{m}\left(1+2(\gamma+\psi(m))\right)+k d^2\right)\\
     &\le&\displaystyle \frac{k}{4}\frac{m!}{d^m}\left(1+2 k d^2\right).
 \end{array}
\end{equation}
Induction step: 
\begin{equation}
 \begin{array}{rcl}
  |d_{n+1}(t)|&\le&\displaystyle\frac 1 2 \left(\int_o^t du |\psi(u)||d_n(u)|+|d_n^{(1)}(t)|+|\int_o^t du \alpha(u) d_n^{(1)}(u)|\right)\\
   &\le&\displaystyle \frac 1 2\left(kk_2 \frac{C_n}{2^n}\frac{(n-2)!}{d^{n-2}}+k\frac{C_n}{2^n}\frac{(n-1)!}{d^{n-1}}+Ak\frac{C_n}{2^n}\frac{(n-2)!}{d^{n-2}}\right)\\
   &\le&\displaystyle k \frac{C_n}{2^{n-1}}\frac{(n-1)!}{d^{n-1}}\left(1+2 k \frac{d^2}{n-1}\right).
 \end{array}
\end{equation}
Moreover, using
\begin{equation}
 \begin{array}{rcl}
    \displaystyle \frac 1 2 |\frac{d^m}{dt^m} \alpha(t) d_n(t)|
   &\le&\displaystyle\frac 1 2 \sum_{l=0}^{m-1}\left(m-1\atop l\right)|\alpha^{(m-l-1)}(t)||d_n^{(l+1)}(t)|\\
   &\le&\displaystyle\frac k{2^{n+1}}\frac{C_n}{d^{n+m-3}}\frac{(m+n-1)!}{n}\\
   &\le&\displaystyle\frac{k}{2^{n+1}}\frac{(m+n+1)!}{d^{m+n-3}}\frac{C_n}{n-1}
 \end{array}
\end{equation}
as well as
\begin{equation}
 \begin{array}{rcl}
   \displaystyle
   \frac 1 2 \left(|\frac{d^{m-1}}{dt^{m-1}}\psi(t) d_n(t)|+|d^{(m+1)}_n(t)|\right)
   \le \displaystyle\frac{k}{2^{n+1}}\frac{(m+n-1)!}{d^{m+n-1}}C_n\left(1+k\frac{d^2}{n-1}\right),
 \end{array}
\end{equation}
we find
\begin{equation}
  |d_{n+1}^{(m)}(t)|\le \frac k{2^{n+1}}\frac{(m+n-1)}{d^{m+n-1}}C_n\left(1+2 k\frac{d^2}{n-1}\right).
\end{equation}
$\blacksquare$\\
If we assume these analyticity conditions and that the initial conditions are such that all integration constants $C_r^{(\l)}$ in the Liouville-Green recursion relations (\ref{final A recurrence}) vanish, we can use equation (\ref{bound on truncation error}), to provide a bound on the truncation error that is independent of the Liouville-Green coefficient. This is rather useful, because the explicit form, particularly of higher order coefficients, is very complicated. We then get
\begin{equation}
 \begin{array}{rcl}
   |\Delta^{(l)}_n(\kappa, t)|
   &\le&\displaystyle \frac{2}{|\kappa|^n}e^{\int_o^t ds|\alpha(s)|}|\dot d^\pm_n(t)| \\
   &\le& \displaystyle\frac{2}{|\kappa|^n}e^{\int_o^t ds|\alpha(s)|}C_n k\frac{(n-1)!}{2^n d^{n-1}}.
  \end{array}
\end{equation}
Insertion into the truncation error yields the bounded
\begin{equation}
 |\epsilon^{(l)}(\kappa,t)|\le\frac{e^{\int_o^t ds |\alpha(s)|}}{\lambda(\kappa,t)}\exp\left[\frac{\int_o^t ds |\chi(s)|}{\lambda(\kappa,t)}\right]\frac{2}{|\kappa|^n}\exp\Big[{\int_o^t ds|\alpha(s)|}\Big]C_n k\frac{(n-1)!}{2^n d^{n-1}}|t|.
\end{equation}
\subsection{\label{RAL}Real Analytic Edges}

We assume that components of the connection and of the edge are real analytic in the chosen trivialization. For $A\ne 0$, we perform a position-independent gauge transformation $\tau_i\mapsto g_o^{-1}\tau_i g_o$, such that $m(0)\ne 0$. Then there are $t_o,\epsilon >0$ s.t. $\inf_{0\ge \tau\ge t_o} \{|m(s)|\}\ge \epsilon$. Specifically, we assume that there exists $t_1>0$ such that for $0\le\tau\le t_1$
\begin{equation}\label{equ:analyticity-condition}
  \begin{array}{rcl}
    \beta_R(\tau)&=&Re(\frac 1 4 M^2(\tau)-\frac 1 2 \dot M(\tau))\\
    \beta_I(\tau)&=&Im(\frac 1 4 M^2(\tau)-\frac 1 2 \dot M(\tau))\\
    \alpha_R(\tau)&=&Re(\dot n(\tau)-M(\tau)n(\tau))\\
    \alpha_I(\tau)&=&Im(\dot n(\tau)-M(\tau)n(\tau))
  \end{array}
\end{equation}
are real analytic for $0\ge\tau\ge t_1$. Then for each $i\in \mathbb N$ there exist
\begin{equation}
  \begin{array}{rclcrcl}
    \alpha_R^i,t_{\alpha R}^i&>&0&\textrm{s.t.}&\alpha_R^i&\ge\sup_{0\le\tau\le t_{\alpha R}^i}\{|\frac{\partial^i\alpha_R(\tau)}{\partial \tau^i}|\}\\
    \alpha_I^i,t_{\alpha I}^i&>&0&\textrm{s.t.}&\alpha_I^i&\ge\sup_{0\le\tau\le t_{\alpha I}^i}\{|\frac{\partial^i\alpha_I(\tau)}{\partial \tau^i}|\}\\
    \beta_R^i,t_{\beta R}^i&>&0&\textrm{s.t.}&\beta_R^i&\ge\sup_{0\le\tau\le t_{\beta R}^i}\{|\frac{\partial^i\beta_R(\tau)}{\partial \tau^i}|\}\\
    \beta_I^i,t_{\beta I}^i&>&0&\textrm{s.t.}&\beta_I^i&\ge\sup_{0\le\tau\le t_{\beta I}^i}\{|\frac{\partial^i\beta_I(\tau)}{\partial \tau^i}|\},
  \end{array}
\end{equation}
which lets us define $\alpha_i:=\max\{\alpha^i_R,\alpha^i_I\}, \beta_i:=\max\{\beta^i_R,\beta^i_I\}$ and
\begin{equation}\label{generalized t_k}
  t_k:=\min\{t_{\alpha R}^o,t_{\alpha I}^o,t_{\beta R}^o,t_{\beta I}^o,...,t_{\alpha R}^k,t_{\alpha I}^k,t_{\beta R}^k,t_{\beta I}^k,t_o,t_1\}.
\end{equation}
We thus have the bounds
\begin{equation}
  \begin{array}{rclcrcl}
    |\alpha(t)|&\le&\alpha_o& \textrm{ for } &t&<&t_o\\
    |\beta(t)|&\le&\beta_o& \textrm{ for } &t&<&t_o,
  \end{array}
\end{equation}
which implies\footnote{Again, $\lambda(\kappa,t):=\inf_{0\le\tau\le t}\{|\kappa+\frac 1 2 \alpha(\tau)|\}$.}
\begin{equation}
  \begin{array}{rcl}
    |\chi(\kappa,t)|
    &=&|\beta(t)-\gamma(\kappa,t)|\le|\beta(t)|+|\gamma(\kappa,t)|
    \\
    &\le& \displaystyle
    \beta_o+\left|\frac{\alpha^2(t)}{4}
           +\frac{3\dot \alpha^2(t)}{16(\kappa+\frac{\alpha(t)}{2})^2}
           -\frac{\ddot \alpha(t)}{4(\kappa+\frac{\alpha(t)}{2})}\right| 
           \textrm{ for } t < t_o
    \\
    &\le&\displaystyle
    \beta_o+\frac 1 4 \alpha_o^2
           +\frac 3{16}\frac{\alpha_1^2}{\lambda^2(\kappa,t)}
           +\frac{\alpha_2}{4\lambda(\kappa,t)} 
           ~~~~~~~~~~~~~\textrm{ for }t<t_2.
  \end{array}
\end{equation}
\begin{equation}
  \begin{array}{rcl}
    \big|\Delta^{(\l)}_1(\kappa,t)\big|
    &=&
    2\left|\mb{e}^{\frac \l 2 \int_0^t \alpha(\tau)\,d\tau}~\dot A_1^{(\l)}(t)\right|
    \\
    &\stackrel{(\ref{dot A^l_n})}{=}&
    2\left|\mb{e}^{\frac \l 2 \int_0^t \alpha(\tau)\,d\tau}~
          \Big(\beta(t)-\frac{1}{4}\alpha^2(t)-\frac{\l}{2}\dot{\alpha}(t)\Big)\right|
    \\
    &\le&
    2\,\mb{e}^{\frac 1 2\int^t_0 d\tau|\alpha(\tau)|}\left|\beta(t)
         -\frac 1 4 \alpha^2(t)
         +\frac{\l}{2} \dot \alpha(t)\right|
    \\
    &\le&
    2\,\mb{e}^{\frac{\alpha_o}2 t}\left(\beta_o+\frac 1 4 \alpha_o^2+\frac 1 2 \alpha_1\right) \textrm{ for }t<t_1.
  \end{array}
\end{equation}
\begin{equation}
  \begin{array}{rcl}
    \big|\phi^{(\l)}_1(\kappa,t)\big|
    &=&\displaystyle
    \int_0^t  \big|\Delta_1^{(\l)}(\kappa,\tau)\big| d\tau
    \\
    &\le& \displaystyle
    \frac 2{\alpha_o}\left(\beta_o+\frac 1 4 \alpha_o^2+\frac 1 2 \alpha_1\right)
    \Big(\mb{e}^{\frac 1 2 \alpha_o t}-1\Big) \textrm{ for }t<t_1.
  \end{array}
\end{equation}
We thus have a bound on the truncation error
\begin{equation}
  \big|\epsilon^{(\l)}_1(\kappa,t)\big|
  \le 
  \mb{e}^{\alpha_o t}
  \Big(\mb{e}^{\frac 1 2 \alpha_o t}-1\Big) \frac{4\beta_o+\alpha_o^2+ 2\alpha_1}{2 \alpha_o \lambda(\kappa,t)}
  \exp\left[\frac{\mb{e}^{\frac 1 2 \alpha_o t}\cdot\big(16\beta_o\lambda^2(\kappa,t)+
                4\alpha_o^2\lambda^2(\kappa,t)+
                3\alpha_1^2+4\alpha_2\lambda(\kappa,t)\big)}
                 {\lambda^3(\kappa,t)}\cdot t\right] ~,
\end{equation}
valid for $t<t_2$ which, using that by construction $\lambda=\mathcal O(|c|)|$ for $c\to \infty\equiv |\kappa|\to \infty$, establishes
\begin{equation}
  \big|\epsilon^{(\l)}_1(\kappa,t)\big|\to 0  \textrm{ for }|c|\to \infty.
\end{equation}

\section{\label{Perturbation Approach}Perturbation Approach}
Restricting to a particular type of edge refers to a background. Even worse it prevents LQC to be continuously embedded into LQG. For this, one would need the solution to (\ref{ODE 2nd order d}) for the isotropic case but for arbitrary edges.

In this section we are now going to construct the general solution to (\ref{ODE 2nd order d}) in terms of a perturbation around a known explicit solution of section \ref{Non-Embedability of Configuration Spaces}. For this we use methods of \cite{Olver}, similar to the Liouville-Green approximation described there. However instead of giving an approximate solution we will compute the {\it exact} solution in terms of a formal power series in the perturbation.

\subsection{\label{General Perturbation}General Perturbation}
Suppose a solution to (\ref{ODE 2nd order d}) is given for a special edge $e_0(t)=\big({e_0}^1(t),{e_0}^2(t),{e_0}^3(t)\big)=:\big(x_0,y_0,z_0\big)$. Now assume the edge $e_0$ is deformed into another edge $e$  such that
$e(t)=\big(x_0+\varepsilon\, \WT{x},~y_0+\varepsilon\, \WT{y},~z_0+\varepsilon\, \WT{z}\big)$, where $ \WT{x}= \WT{x}(t),~\WT{y}= \WT{y}(t),~\WT{z}= \WT{z}(t),  $ and  $\varepsilon=const$ is a deformation parameter. Then we can write the components of the tangent vector ${\bf \dot{e}}$ as

\be
  \dot x = \dot x_0 +\e~ \dot{\WT{x}} ~~~~~~~~
  \dot y = \dot y_0 +\e~ \dot{\WT{y}} ~~~~~~~~
  \dot z = \dot z_0 +\e~ \dot{\WT{z}} ~~~.
  \nonumber
\ee
This implies
\barr{llcclccllcccc}\label{m with perturbation}
  m&=&(\dot x_0 -\mb{i}\dot y_0)&+&\e&(\dot{\WT{x}}-\mb{i}\dot{\WT{y}})
  &~~~~~~~~~&
  n&=&\dot z_0 &+&\e&\dot{\WT{z}}_0
  \\[-2mm]
  &=:&m_0 &+&\e&\WT{m}
  &~~~&
  &=:& n_0&+&\e&\WT n
\earr
Moreover we introduce
\barr{lclclclll}
   M_0:=\frac{\dot m_0}{m_0}
   &~~~\text{and}~~~~&
   \WT{M}:=\frac{\WT{m}}{m_0}
   &~~~\text{with}~&
   &\dot{\WT{M}}&=&\frac{\dot{\WT{m}}}{m_0}-\WT{M}M_0
   \NN
   &&&&&\ddot{\WT{M}}&=&\frac{\ddot{\WT{m}}}{m_0}-2M_0\big(\dot{\WT{M}}+\WT{M}M_0 \big)
   \nonumber
\earr
in order to expand
\ba
  M&=&\frac{\dot m}{m}
    =\frac{\dot m_0 +\e \dot{\WT{m}}}{m_0 + \e \WT m}
    =\frac{1}{m_0}\big(\dot m_0 +\e \dot{\WT{m}}\big)\frac{1}{1-(-\e\WT{M})}
    \NN
    &=&\big(M_0+\e(\dot{\WT{M}}+\WT M M_0)\big)~\sum_{k=0}^{\infty}(-1)^k\e^k \WT{M}^k
    \NN
    &=&\sum_{k=0}^\infty (-1)^k\Big\{\e^k M_0\WT M^k + \e^{k+1}(\dot{\WT{M}}\WT M^k +M_0\WT M^{k+1})\Big\}
    \NN
    &=&M_0 +\sum_{k=1}^{\infty}(-1)^k\e^k\Big\{\cancel{M_0\MT^k}-\dot\MT \MT^{k-1}-\cancel{M_0\MT^k}\Big\}
    \NN
    &=& M_0 -\frac{\dot\MT}{\MT}\sum_{k=1}^\infty (-1)^k \e^k \MT^k
    \nonumber
\ea
where we need to assume $|\e\WT M|<1$ from the first to second line in order to ensure that the geometric series converges. Notice that this assumption is  re-parametrization independent: If we introduce a new edge parameter $T=T(t)$, then for an arbitrary function $f(t)$ it holds that $\dot{f}(t)=\frac{d f(t)}{dt}=\frac{df(T)}{dT} \dot{T}(t)$, hence the $\dot{T}$-terms cancel by construction of $\WT{M}$. As a consequence, it is straight forward to write down

\ba
   M^2&=& M_0^2
         -2 M_0\frac{\dot \MT}{\MT}\sum_{k=1}^\infty(-1)^k\e^k\MT^k
         +\bigg[\frac{\dot\MT}{\MT}\bigg]^2
         \sum_{k=1}^\infty \sum_{l=1}^\infty (-1)^{k+l}\e^{k+l}\MT^{k+l}
      \NN
      &=& M_0^2
         -2 M_0\frac{\dot \MT}{\MT}\sum_{k=1}^\infty(-1)^k\e^k\MT^k
         +\bigg[\frac{\dot\MT}{\MT}\bigg]^2
          \sum_{r=2}^\infty(r-1)(-1)^r\e^r\MT^r
      \NN
      &=& M_0^2 + \big(2 M_0\dot\MT \big)~\e
                + \frac{\dot \MT}{\MT}\sum_{r=2}^\infty(-1)^r\e^r\MT^r \bigg\{(r-1)\frac{\dot\MT}{\MT}-2M_0\bigg\}
  \nonumber
\ea
and
\ba
   \dot M&=&\dot M_0 -\frac{\ddot \MT \MT -\dot\MT^2}{\MT^2}
            \sum_{k=1}^\infty (-1)^k \e^k \MT^k
            -\frac{\dot \MT}{\MT} \sum_{k=1}^\infty (-1)^k \e^k~ k~ \MT^{k-1}\dot\MT
         \NN
         &=&\dot M_0-\bigg(\frac{\ddot\MT}{\MT}-\bigg[\frac{\dot\MT}{\MT}\bigg]^2\bigg)
         \sum_{k=1}^\infty (-1)^k \e^k \MT^k
         - \bigg[\frac{\dot\MT}{\MT}\bigg]^2 \sum_{k=1}^\infty (-1)^k \e^k~k~ \MT^k
         \NN
         &=& \dot M_0 +\ddot\MT~\e-\MT^{-2}\sum_{k=2}^\infty (-1)^k~ \e^k~ \MT^k
            \big\{(k-1){\dot{\MT}}^2+\ddot\MT\MT\big\}~~~~.
   \nonumber
\ea
Finally we use the expansion
\barr{llcccccccccc}
  |m|^2 +n^2&=& \multicolumn{7}{l}{  ( \dot x_0 +\e~\dot{\WT{x}})^2
                                    +(\dot y_0 +\e~\dot{\WT{y}})^2
                                    +(\dot z_0 +\e~\dot{\WT{z}})^2                         }
            \\
            &=&   {\dot x_0}^2 + \dot y_0^2 + \dot z_0^2
               &+&2\big(\dot x_0\dot{\WT{x}}+\dot y_0\dot{\WT{y}}+\dot z_0\dot{\WT{z}}\big)&\e&
               &+&\big(\dot{\WT x}^2+\dot{\WT y}^2+\dot{\WT z}^2\big)&\e^2&
            \\
            &=:&    |m_0|^2+n_0^2
               &+& 2 Y &\e&
               &+& \big(|\WT{m}|^2+\WT{n}^2\big)&\e^2&        ~~~,
    \nonumber
\earr
in order to express

\ba
  N&=&\mb{i}c\big(\dot{n}-Mn+\mb{i}c(m\cc m +n^2)\big)
  \NN
  &=&-c^2\Big(|m_0|^2+n_0^2
               + 2 Y \e
               + \big(|\WT{m}|^2+\WT{n}^2\big)\e^2\Big)
     +\mb{i}c\big(\dot n_0+\e~\dot{\WT{n}}\big)
     -\mb{i}c\big( n_0+\e~{\WT{n}}\big)~M
  \NN
  &=&-c^2\Big(\ldots\Big)
     +\mb{i}c\big(\dot n_0+\e~\dot{\WT{n}}\big)
     -\mb{i}c\big( n_0+\e~{\WT{n}}\big)~
     \Big\{M_0 -\frac{\dot\MT}{\MT}\sum_{k=1}^\infty (-1)^k \e^k \MT^k\Big\}
  \NN
  &=&-c^2\Big(\ldots\Big)
     +\mb{i}c\big(\dot n_0 +\e~\dot{\WT{n}}-M_0(n_0+\e\WT{n})\big)
     +\mb{i}c\frac{\dot\MT}{\MT}\sum_{k=1}^\infty (-1)^k\MT^k\big\{n_0~\e^k+\WT{n}~\e^{k+1}\big\}
  \NN
  &=&-c^2\Big(\ldots\Big)
     +\mb{i}c\big(\dot n_0 +\e~\dot{\WT{n}}-M_0(n_0+\e\WT{n})\big)
     +\mb{i}c\Big(-n_0\dot\MT~\e +\frac{\dot\MT}{\MT}\Big(n_0-\frac{\WT{n}}{\MT}\Big)
     \sum_{k=2}^\infty (-1)^k\MT^k~\e^k\Big)
  \NN\NN
  &=& -c^2\big(|m_0|^2+n_0^2\big)+\mb{i}c \big(\dot n_0 -M_0~ n_0\big)
      +\big\{-2c^2Y+\mb{i}c\big(\dot{\WT{n}}-M_0\WT{n}-n_0\dot{\WT{M}}\big)\big\}~\e
  \NN
  && +\big\{-c^2 (|\WT{m}|^2+\WT{m}^2) +\mb{i}c~ \dot\MT\big(n_0 \MT -\WT{n}\big)     \big\}~\e^2
     +\mb{i}c~\frac{\dot\MT}{\MT}\Big(n_0-\frac{\WT{n}}{\MT}\Big)
     \sum_{k=3}^\infty (-1)^k\MT^k~\e^k
   \nonumber
\ea
So we are finally able to expand the bracket term in (\ref{ODE 2nd order d}) into orders of $\e$ :
\ba\label{Full perturbation expansion of rhs}
  \Big\{\frac{1}{4} M^2-\frac{1}{2}\dot{M} +N\Big\}
  &=&\frac{1}{4} M_0^2-\frac{1}{2}\dot M_0 -c^2\big(|m_0|^2+n_0^2\big)+\mb{i}c \big(\dot n_0 -M_0~ n_0\big)
  \NN
  &&+\e~\Big\{\frac{1}{2}\big(M_0\dot\MT-\ddot\MT\big)
     -2c^2Y+\mb{i}c\big(\dot{\WT{n}}-M_0\WT{n}-n_0\dot{\WT{M}}\big)\Big\}
  \NN
  &&+\e^2~\Big\{\frac{1}{4}\Big(({\dot{\MT}})^2-2 M_0\MT\dot{\MT}\Big)
               +\frac{1}{2}\Big((\dot\MT)^2+\ddot\MT\MT\Big)
               -c^2 (|\WT{m}|^2+\WT{n}^2) +\mb{i}c~ \dot\MT\big(n_0 \MT -\WT{n}\big) \Big\}
  \NN
  &&+\sum_{k=3}^\infty(-1)^k\MT^k~\e^k~
     \Big\{\frac{1}{4}\dot{\MT}\MT^{-1}\Big((k-1)\dot\MT\MT^{-1}-2M_0\Big)
          +\frac{1}{2}\MT^{-2}\Big((k-1)({\dot{\MT}})^2+\ddot\MT\MT\Big)
  \NN&&~~~~~~~~~~~~~~~~~~~~~~~~~~~~~~
          +\mb{i}c~\dot\MT\MT^{-1}\Big(n_0-\WT{n}\MT^{-1}\Big)\Big\}   ~~~~,
\ea
where we have assumed $|\e\WT M|<1$ and $Y:=\dot x_0\dot{\WT{x}}+\dot y_0\dot{\WT{y}}+\dot z_0\dot{\WT{z}}$.

\subsection{\label{Simplifications}Simplifications}

As it stands (\ref{Full perturbation expansion of rhs}) holds in full generality. However, in order to solve  (\ref{ODE 2nd order d}) with the perturbation Ansatz, we require $M_0$ and $n_0$ to be constants. Hence $\dot{M}_0=0=\dot{n}_0$. Moreover we may assume that we perturb $e_0(t)$ only in perpendicular direction, that is for the edge tangents $\dot{\bf e}_0, \dot{\bf e}$ it holds that $\big<\dot{\bf e}_0,\dot{\bf e}\big>=\big<\dot{\bf e}_0,\dot{\bf e}_0\big>=|\dot{\bf e}_0|^2$. Then it follows that $Y:=\dot x_0\dot{\WT{x}}+\dot y_0\dot{\WT{y}}+\dot z_0\dot{\WT{z}}=0$.
\\
~\\
If we leave the beginning point of the edge fixed under perturbation we have $e_0(t=0)=e(t=0)$. Moreover without loss of generality we can choose $e_0(t)$ such that $\dot{\bf e}_0(t=0)=\dot{\bf e}(t=0)$.
\\
~\\
Finally we are free to choose a convenient parametrization, for example arc-length parametrization of $e_0(t)$, such that $|m_0|^2+n_0^2 =1$. We will refrain from the latter and write using \fbox{$\kappa:=\mb{i}\, c$}
\ba\label{Full perturbation expansion of rhs simplified}
  \Big\{\frac{1}{4} M^2-\frac{1}{2}\dot{M} +N\Big\}
  &=&\frac{1}{4} M_0^2 +\kappa^2\big(|m_0|^2+n_0^2\big)-\kappa\, M_0~ n_0
  \NN
  &=&+\e~\Big\{\frac{1}{2}\big(M_0\dot\MT-\ddot\MT\big)
     +\kappa\big(\dot{\WT{n}}-M_0\WT{n}-n_0\dot{\WT{M}}\big)\Big\}
  \NN
  &&+\e^2~\Big\{\frac{1}{4}\Big(({\dot{\MT}})^2-2 M_0\MT\dot{\MT}\Big)
               +\frac{1}{2}\Big((\dot\MT)^2+\ddot\MT\MT\Big)
               +\kappa^2 (|\WT{m}|^2+\WT{n}^2) +\kappa~ \dot\MT\big(n_0 \MT -\WT{n}\big) \Big\}
  \NN
  &&+\sum_{k=3}^\infty(-1)^k\MT^k~\e^k~
     \Big\{\frac{1}{4}\dot{\MT}\MT^{-1}\Big((k-1)\dot\MT\MT^{-1}-2M_0\Big)
          +\frac{1}{2}\MT^{-2}\Big((k-1)({\dot{\MT}})^2+\ddot\MT\MT\Big)
  \NN&&~~~~~~~~~~~~~~~~~~~~~~~~~~~~~~
          +\kappa~\dot\MT\MT^{-1}\Big(n_0-\WT{n}\MT^{-1}\Big)\Big\}
  \NN
  &=:&\K^2(\kappa) +\sum_{k=1}^\infty \e^k f_k(\kappa,t)~~~,
\ea
where we have introduced the obvious shorthand in the last line. In particular
\be\label{definition K}
   \K^2(\kappa):=\frac{1}{4} M_0^2 +\kappa^2\big(|m_0|^2+n_0^2\big)-\kappa\, M_0~ n_0
\ee
\subsection{Solution by the Liouville-Green Method}
With the expansion (\ref{Full perturbation expansion of rhs simplified})  we can write (\ref{ODE 2nd order d}) as
\be\label{Start LG}
   \ddot{d}=\Big\{\K^2(\kappa)+ \sum_{k=1}^\infty \e^k~f_{k}(\kappa,t)\Big\}~d
\ee
Now we make a Liouville-Green Ansatz, similar to section \ref{LG}. That is we set $d=d(\kappa,t)=\displaystyle\sum_{\sigma=\pm}d^{(\sigma)}(\kappa,t)$ with
\ba\label{Ansatz LG}
   d(\kappa,t)=\sum_{\sigma=\pm}d^{(\sigma)}(\kappa,t)
   &:=&\sum_{\sigma=\pm}\mb{e}^{\sigma\K t}\sum_{n=0}^\infty d_n^{(\sigma)}(\kappa,t) \cdot \varepsilon^n
   \NN
   \dot{d}(\kappa,t)=\sum_{\sigma=\pm}\dot{d}^{(\sigma)}(\kappa,t)
   &:=&\sum_{\sigma=\pm}\mb{e}^{\sigma\K t}\sum_{n=0}^\infty \Big(\sigma\,\K\, d_n^{(\sigma)}(\kappa,t)+\dot{d}_n^{(\sigma)}(\kappa,t) \Big)\cdot \varepsilon^n
   \NN
   \ddot{d}(\kappa,t)=\sum_{\sigma=\pm}\ddot{d}^{(\sigma)}(\kappa,t)
   &:=&\sum_{\sigma=\pm}\mb{e}^{\sigma\K t}\sum_{n=0}^\infty \Big(\K^2\, d_n^{(\sigma)}(\kappa,t)+2\,\sigma\,\K\,\dot{d}_n^{(\sigma)}(\kappa,t)
   + \ddot{d}_n^{(\sigma)}(\kappa,t)\Big)\cdot \varepsilon^n
\ea
In the last line we have used the fact that $\sigma^2=1$. As we will see at the end of this computation the \linebreak {\bf choice of initial conditions}
\be\label{initial conditions LG-Ansatz}
   d^{(\sigma)}(\kappa,0)=1~~~~\text{and}~~~~\dot{d}^{(\sigma)}(\kappa,0)=0~~~,
\ee
which have to hold for {\it arbitrary} $\varepsilon$, will be convenient. Now we plug the Ansatz (\ref{Ansatz LG}) into (\ref{Start LG}). For clarity we will again suppress the dependence on $t,\kappa$ and simply write e.g. $d_n^{(\sigma)}$ instead of $d_n^{(\sigma)}(\kappa,t)$. This gives:
\[
   \sum_{\sigma=\pm}\mb{e}^{\sigma\K t}\sum_{n=0}^\infty
   \Big\{
      \cancel{\K^2\,d_n^{(\sigma)}}
             +2\,\sigma\,\K\,\dot{d}_n^{(\sigma)}
             + \ddot{d}_n^{(\sigma)}
             - \cancel{\K^2\,d_n^{(\sigma)}}
             - d_n^{(\sigma)}\sum_{k=1}^\infty f_k \varepsilon^k
   \Big\}\varepsilon^n=0
\]
This has to hold at {\it any point of} $e(t)$, that is for {\it arbitrary} values of $t$. Hence, it must hold that
\[
   \sum_{n=0}^\infty
   \Big\{
             2\,\sigma\,\K\,\dot{d}_n^{(\sigma)}
             + \ddot{d}_n^{(\sigma)}
             - d_n^{(\sigma)}\sum_{k=1}^\infty f_k\, \varepsilon^k
   \Big\}\varepsilon^n
   =\sum_{n=1}^\infty \varepsilon^n
   \Big\{
               \ddot{d}_n^{(\sigma)}
             + 2\,\sigma\,\K\,\dot{d}_n^{(\sigma)}
             - \sum_{k=1}^n d_{n-k}^{(\sigma)} f_k
   \Big\}
             + \ddot{d}_0^{(\sigma)} + 2\,\sigma\,\K\,\dot{d}_0^{(\sigma)}
   =0
\]
This has to hold for arbitrary values of $\varepsilon$, hence separately in {\it every} order $n$ of $\varepsilon$:
\ba
  \fbox{$\MC{O}(\varepsilon^0)$}&~~&\ddot{d}_0^{(\sigma)} + 2\,\sigma\,\K\,\dot{d}_0^{(\sigma)} = 0
  \\\label{C2}
  \fbox{$\MC{O}(\varepsilon^{n>0})$}&&\displaystyle \ddot{d}_n^{(\sigma)}
             + 2\,\sigma\,\K\,\dot{d}_n^{(\sigma)}
             = \lambda_n^{(\sigma)}
             ~~~~\text{with}~~\lambda_n^{(\sigma)}:=\sum_{k=1}^n d_{n-k}^{(\sigma)}\, f_k
\ea
\subsubsection{Solution to $\MC{O}(\varepsilon^0)$:} This is a homogeneous linear ODE of second order with constant coefficients. Its solution is given by
\[
   d_0^{(\sigma)}=A_0^{(\sigma)}+B_0^{(\sigma)}\,\mb{e}^{-2\sigma\K t}
\]
with integration constants $A_0^{(\sigma)}, B_0^{(\sigma)}$.

\subsubsection{Solution to $\MC{O}(\varepsilon^{n>0})$:} This second order ODE is inhomogeneous but still linear with constant coefficients. Its general solution  $d_n^{(\sigma)}$ can be obtained as a linear combination
\be\label{special solution}
   d_n^{(\sigma)}=d_{n,\text{HOM}}^{(\sigma)}+ d_{n,\text{SP}}^{(\sigma)}
\ee
of the general solution $d_{n,\text{HOM}}^{(\sigma)}$ to the homogeneous equation plus a special solution  $d_{n,\text{SP}}^{(\sigma)}$ to the inhomogeneous equation. The homogeneous part is equivalent to the $\MC{O}(\varepsilon^0)$-case and given by
\[
   d_{n,\text{HOM}}^{(\sigma)}=A_{n,\text{HOM}}^{(\sigma)}+B_{n,\text{HOM}}^{(\sigma)}\,\mb{e}^{-2\sigma\K t}
\]
with integration constants $A_{n,\text{HOM}}^{(\sigma)}, B_{n,\text{HOM}}^{(\sigma)}$. The special solution $d_{n,\text{SP}}^{(\sigma)}$ can be obtained from $d_{n,\text{HOM}}^{(\sigma)}$ by the method of variation of constants, that is we make the Ansatz
\[
   d_{n,\text{SP}}^{(\sigma)}(\kappa,t)
   =A_n^{(\sigma)}(\kappa,t)+B_n^{(\sigma)}(\kappa,t)\,\mb{e}^{-2\sigma\K t}~~~.
\]
With the usual requirement: \fbox{$I$} $\dot{A}_n^{(\sigma)}(\kappa,t)+\dot{B}_n^{(\sigma)}(\kappa,t)\,\mb{e}^{-2\sigma\K t}
   =0$ this gives
\barr{lcl}
   \dot{d}_{n,\text{SP}}^{(\sigma)}(\kappa,t)
   &=&-2\sigma\,\K\, B_n^{(\sigma)}(\kappa,t)\,\mb{e}^{-2\sigma\K t}
   \\
   \ddot{d}_{n,\text{SP}}^{(\sigma)}(\kappa,t)
   &=&\big(-2\sigma\,\K\, \dot{B}_n^{(\sigma)}(\kappa,t)
           +4\,\K^2B_n^{(\sigma)}(\kappa,t) \big)\mb{e}^{-2\sigma\K t}
   \nonumber
\earr
If we plug this Ansatz into (\ref{C2}) we get (recall that $\sigma^2=1$): \fbox{$II$} $\dot{B}_n^{(\sigma)}(\kappa,t)
=-\frac{1}{2\,\sigma\,\K}\,\mb{e}^{2\sigma\K t}\,\lambda_n^{(\sigma)}(\kappa,t)$ and thus in turn
\ba\nonumber
    B_n^{(\sigma)}(\kappa,t)
    &=&-\frac{\sigma}{2\,\K}\int_0^t \mb{e}^{2\sigma\K s}\,\lambda_n^{(\sigma)}(\kappa,s)~ds + B_{n,\text{SP}}^{(\sigma)}
    \\\nonumber
    A_n^{(\sigma)}(\kappa,t)
    &=&\displaystyle\frac{\sigma}{2\,\K}\int_0^t \,\lambda_n^{(\sigma)}(\kappa,s)~ds + A_{n,\text{SP}}^{(\sigma)}
    \nonumber
\ea
where $A_{n,\text{SP}}^{(\sigma)},B_{n,\text{SP}}^{(\sigma)}$ are again integration constants. Therefore the special solution $d_{n,\text{SP}}^{(\sigma)}(\kappa,t)$ is given by
\be\nonumber
  d_{n,\text{SP}}^{(\sigma)}(\kappa,t)
  =A_{n,\text{SP}}^{(\sigma)}+B_{n,\text{SP}}^{(\sigma)}
   +\frac{\sigma}{2\,\K}
   \int_0^t \big(1- \mb{e}^{2\sigma\,\K\,(s-t)} \big)\,\lambda_n^{(\sigma)}(\kappa,s)~ds
\ee
Using (\ref{special solution}) the final solution $d_n^{(\sigma)}$ of (\ref{C2}) is given by
\be\label{solution dn_sigma}
   d_n^{(\sigma)}
   = D_{n,\text{SP}}^{(\sigma)}
   + B_{n,\text{HOM}}^{(\sigma)}\,\mb{e}^{-2\sigma\K t}
   +\frac{\sigma}{2\,\K}
   \int_0^t \big(1- \mb{e}^{2\sigma\,\K\,(s-t)} \big)\,\lambda_n^{(\sigma)}(\kappa,s)~ds
\ee
where we have introduced the overall integration constant
$D_{n,\text{SP}}^{(\sigma)}:=A_{n,\text{HOM}}^{(\sigma)}+A_{n,\text{SP}}^{(\sigma)}+B_{n,\text{SP}}^{(\sigma)}$ .

\subsubsection{Implementation of Initial Conditions}

Now we are going to implement the initial conditions (\ref{initial conditions LG-Ansatz}) into (\ref{solution dn_sigma}).
\ba\nonumber
   d^{(\sigma)}(\kappa,0)
   &=&1
     =\sum_{n=0}^\infty d_n^{(\sigma)}(\kappa,0)\cdot\varepsilon^n
     =A_0^{(\sigma)}+B_0^{(\sigma)}
       + \sum_{n=1}^\infty \big(D_{n,\text{SP}}^{(\sigma)}+B_{n.\text{HOM}}^{(\sigma)}\big)\cdot\varepsilon^n
   \NN
   \dot{d}^{(\sigma)}(\kappa,0)
   &=&0
     =\sum_{n=0}^\infty d_n^{(\sigma)}(\kappa,0)\cdot\varepsilon^n
     =-2\sigma\,\K\Big( B_0^{(\sigma)}+ \sum_{n=1}^\infty B_{n,\text{HOM}}^{(\sigma)}\cdot\varepsilon^n\Big)  \nonumber
\ea
As these conditions have to hold for {\it arbitrary $\varepsilon$}, it follows that
\be\nonumber
   A_0^{(\sigma)}=1~~~~
   B_0^{(\sigma)}=0~~~~~~~~\text{and}~~~~
   B_{n.\text{HOM}}^{(\sigma)}=0=D_{n,\text{SP}}^{(\sigma)}
\ee
Therefore
\be\label{solution dn_sigma with initial conditions}
   d^{(\sigma)}(\kappa,t)
   = \mb{e}^{\sigma\,\K\, t} \bigg\{
   1+\frac{\sigma}{2\,\K}\sum_{n=1}^\infty\varepsilon^n
   \int_0^t \big(1- \mb{e}^{2\sigma\,\K\,(s-t)} \big)\,\lambda_n^{(\sigma)}(\kappa,s)~ds
   \bigg\}
\ee
where $\displaystyle\lambda_n^{(\sigma)}(\kappa,t):=\sum_{k=1}^n d_{n-k}^{(\sigma)}(\kappa,t)\, f_k(\kappa,t)$. Moreover by construction  $d_0^{(\sigma)}(\kappa,t)=1$ and $f_k(\kappa,t)$ is given according to (\ref{Full perturbation expansion of rhs simplified}) and $\K$ is defined in (\ref{definition K}).

\subsection{Solution}

Now we are set up to construct the solution to (\ref{ODE 2nd order a}) for an arbitrary edge $e(t)$ using the Ansatz (\ref{Ansatz general solution for a}) in terms of a perturbation about an edge $e_0(t)$. According to section \ref{Simplifications} we demand
\[
 e_0(0)=e(0)\text{~~~~and~~~~}  \dot{\bf e}_0(0)=\dot{\bf e}(0)~.
 \]
This implies  $m(0)=m_0(0)$. Imposing the initial conditions  (\ref{ODE 2nd order a initial conditions})  we obtain:
\ba
   a(\kappa,0)=1&=&A_{(+)}d^{(+)}(\kappa,0)+A_{(-)}d^{(-)}(\kappa,0)=\sqrt{m_0(0)}\,\big(A_{(+)} +A_{(-)}\big)
   \NN
   \dot{a}(0)=\kappa n(0)&=&A_{(+)}\dot{d}^{(+)}(\kappa,0)+A_{(-)}\dot{d}^{(-)}(\kappa,0)
             = \K\,(A_{(+)} - A_{(-)})
   \nonumber
\ea
Hence $A_{(\sigma)}=\frac{1}{2}\big[\frac{1}{\sqrt{m_0(0)}}+\sigma \frac{\kappa\,n_0(0)}{\K(\kappa)}\big]$ and we get the final solution
\be\label{final solution for a}
     a(\kappa,t)=\sqrt{m}(t)\sum_{\sigma=\pm}
     A_{(\sigma)}\,
     \mb{e}^{\sigma\,\K\, t} \bigg\{
   1+\frac{\sigma}{2\K}\sum_{n=1}^\infty\varepsilon^n\cdot
   \int_0^t \big(1-\mb{e}^{2\sigma\,\K\,(s-t)}\big) \,\lambda_n^{(\sigma)}(\kappa,s)~ds
   \bigg\}
\ee
with $\lambda_n^{(\sigma)}(\kappa,s)$ and $\K=\K(\kappa)$ given below (\ref{Full perturbation expansion of rhs simplified}) and $m$ given in (\ref{m with perturbation}).

\subsection{\label{Perturbation about a Line}Perturbation about a Line}

Certainly the functions $f_k(\kappa,t)$ of (\ref{Full perturbation expansion of rhs simplified}), which are needed in order to explicitly compute (\ref{final solution for a}), are still quite complicated. Also the definition (\ref{definition K}) of $\K$ involves a square root taken from a complex number.
To simplify this situation we can  construct a solution of (\ref{ODE 2nd order d}) for a general edge $e(t)$ as follows. Given $e(t)$ we construct the solution to (\ref{ODE 2nd order d}) as a perturbation of a line $e_0(t)$, for which again
\be\label{line conditions 1}
   e_0(0)=e(0)~\text{   and   }~ \dot{\bf e}_0(0)=\dot{\bf e}(0)
\ee
holds. W.l.o.g. we can choose the maximal simplification for $e_0$ being a line. That is
\be\label{line conditions 2}
   e_0(t)=\big(x_0(t),0,0\big)~~~
     e(t)=\big(x_0(t)\,,\, \e\,\WT{y}(t)\,,\,\e\,\WT{z}(t)\big)
\ee
Moreover we choose arc-length parametrization of the line, that is we demand
\ba\label{arc length parametrization}
    1&=&\dot{x}_0^2=m(0)=m_0(0)
\ea
Then we have
\barr{lclclclclclclc}
   m_0&=&\dot{x}_0=1 &~~&M_0&=&0=\dot{m}_0 &~~&n_0&=&0=\dot{n}_0
   &~~&~\K^2=\kappa^2=-c^2~~~\WT{m}=\dot{\WT{y}}
   \\
   \WT{m}&=&\dot{\WT{y}} &&\WT{M}&=&\dot{\WT{m}}=\ddot{\WT{y}}
   &&\WT{n}&=&\dot{\WT{z}}
\earr
Under these assumptions, expression  (\ref{Full perturbation expansion of rhs simplified}) can be simplified to
\ba\label{Line perturbation expansion of rhs}
  \Big\{\frac{1}{4} M^2-\frac{1}{2}\dot{M} +N\Big\}
  &=&\kappa^2+\e~\kappa\,\dot{\WT{n}}
             +\e^2~\Big\{\frac{3}{4}({\dot{\MT}})^2
               +\frac{1}{2}\ddot\MT\MT
               +\kappa^2 (|\WT{m}|^2+\WT{n}^2)
               -\kappa~ \dot\MT\WT{n} \Big\}
  \NN
  &&+\sum_{k=3}^\infty\e^k~(-1)^k\,\MT^{k-2}~
     \Big\{\frac{3}{4}(k-1)(\dot{\MT})^2
          +\frac{1}{2}\ddot\MT\MT
          -\kappa~\dot\MT\WT{n}\Big\}
\ea
and expression (\ref{final solution for a}) reads
\be\label{line solution for a}
     a(\kappa,t)=\sum_{\sigma=\pm}
     \mb{e}^{\sigma\,\kappa\, t} \bigg\{
   1+\frac{\sigma}{2\kappa}\sum_{n=1}^\infty\varepsilon^n\,
   \int_0^t \big(1-\mb{e}^{2\sigma\,\kappa\,(s-t)}\big) \,\lambda_n^{(\sigma)}(\kappa,s)~ds
   \bigg\}
\ee
Notice, that if we set \fbox{$\e=c^{-1}$} then for $c\rightarrow\infty$ the solution (\ref{line solution for a}) takes the form
\be
     a(\kappa,t)=\sum_{\sigma=\pm}
     \mb{e}^{\sigma\,\kappa\, t} \Big\{
   1+ \mathcal{O}(c^{-1})  \Big\}~~.
\ee
In fact, this property holds for arbitrary curves in the limit  $c\rightarrow\infty$. The proof is given in section \ref{Asymptotics}.
%
\section{\label{Geometry Ashtekar Vars}Geometric Interpretation of the Parameter $c$}

\subsection{The Ashtekar Connection Revisited}
For completeness we will briefly sketch some of the insights obtained in \cite{Levermann2009}, where a coordinate free treatment of the Ashtekar connection is developed. Notice that unlike in the rest of this paper we denote vectors by the symbol '$e$'. 
~\\~\\
We are working with a (3+1)-decomposition of spacetime $(\M,g)\cong \mb{R}\times\Sigma$. The Cauchy surfaces $\Sigma$ are orientable. Hence at every $m\in \M$ we can decompose the tangent space $T_m\M=T_m\M^\|\oplus T_m\M^\perp$ into two orthogonal subspaces, the component $T_m\M^\|\cong T_m\Sigma$ tangent to $\Sigma$ and the normal component $T_m\M^\perp\cong N n$, where $n$ is the surface surface normal unit vector and $N\in \mb{R}$. With the signature $(s,+,+,+)$ of $\M$ ($s=\pm 1$) we have $g(n,n)=s$ and $g(n,X)$=0, $g(X,X)>0$ for every $X\in T_m\Sigma$. To shorten notation, we use the isomorphism between the local tangent spaces $T_m\Sigma\sim\mb{R}^3$ and the Lie algebra $so(3)$ induced by the equivalence of the defining representation of $SO(3)$ on $\mb{R}^3$ and the adjoint representation of $SO(3)$ on $so(3)$. Let $\{e_I\}_{I=1,2,3}$ be an orthonormal basis-frame of $T_m\Sigma$, e.g. the Cartesian standard basis of $\mb{R}^3$. Moreover let $\{\tau_I\}_{I=1,2,3}$ be a basis of $so(3)\sim su(2)$, orthonormal with respect to the Cartan-Killing-metric thereon, e.g.  $(\tau_I)_{JK}=-\epsilon_{IJK}$. Now we choose a fixed identification $e_I\leftrightarrow \tau_I$ and can write any element $X\in T_m\Sigma$ equivalently as $X=\sum_I X_I e_I \leftrightarrow \sum_I X_I\tau_I$, where $X_I=g(X,e_I)$ denotes the expansion of $X$ into  $\{e_I\}_{I=1,2,3}$. Consequently we will write capital Latin indices to denote both $so(3)$ indices and indices with respect to the orthonormal frame $\{e_I\}$. Then we can decompose the covariant derivative $\nabla^\M$ on $\M$ coming from the Levi-Civita-connection as
\ba
   \nabla^\M_{e_K} e_I 
   &=& (\nabla^\M_{e_K} e_I)^\| + (\nabla^\M_{e_K} e_I)^\perp
   \NN
   &=& \sum_J g(\nabla^\M_{e_K} e_I ,e_J)~e_J +  s~g(\nabla^\M_{e_K} e_I ,n)~n
   \NN
   &=& \sum_J g(\nabla^{\Sigma}_{e_K} e_I ,e_J)~e_J +  s~g(K(e_K,e_I) ,n)~n~~,
\ea
where in the last line we denote by $\nabla^{\Sigma}$ the covariant derivative coming from Levi-Civita-connection on $\Sigma$, which is for any $e_I,e_J\in T_m\Sigma$ precisely given by the tangential component of $\nabla^\M$. Now we introduce the shorthands
\be\label{matrix elements Weingarten map}
   k_{KI}=k(e_K,e_I)
         =g(K(e_K,e_I) ,n)
        :=g(\nabla^\M_{e_K} e_I ,n)
         =-g(e_I,\nabla^\M_{e_K} n)
         =:-g(e_I,W_n(e_K))~~,
\ee
where $k_{KI}$ are the components of extrinsic curvature and $W_n:T_m\Sigma\rightarrow T_m\Sigma$ is called Weingarten map\footnote{By construction $W_n$ has only components in $T_m\Sigma$. We have $g(n,n)=s=const$ and hence $(\nabla^\M_{e_K}g)(n,n)=0=\underbrace{e_K(s)}_{=0}+2g(\nabla^\M_{e_K}n,n)$, where we use the symmetry of $g$ and the fact that $\nabla^\M$ is metric.}. Notice the symmetry of $k_{KI}=\frac{1}{2}(k_{KI}+k_{IK})=k_{(KI)}$: By construction torsion $T^\M$ of the Levi-Civita-Connection $\nabla^\M$  vanishes and we have
\[
   T^\M(e_K,e_I)=0=\nabla^\M_{e_K}e_I - \nabla^\M_{e_I}e_K -[e_K,e_I]
\]
and therefore $k_{KI}=k_{IK}$ if the vectors $e_K,e_I$ commute, that is $[e_K,e_I]=0$.
For the Ashtekar connection we have for any two smooth vector fields $X,Y\in \Gamma(T\Sigma)$:
\be\label{action of nabla^A}
   \nablaA_X Y=\nabla^{\Sigma}_X Y+\beta~ W_n(X)\bullet Y
\ee
where $\beta$ is the Barbero-Immirzi parameter and we have 
introduced the product:
\be\label{induced product in TSigma}
   X\bullet Y=\sum_{I,J}X_IY_J~e_I\bullet e_J=\sum_{I,J,K}X_IY_J~\epsilon_{IJK}~e_K~~,
\ee
where we have expanded $X=\sum_{I}g(X,e_I)~e_I=:\sum_I X_I~ e_I$. We can write this for the orthonormal basis $\{e_I\}_{I=1,2,3}$ as 
\ba\label{nablaA in ONB}
   \nablaA_{e_I} e_J
   &=&\nabla^{\Sigma}_{e_I} e_J-\beta~\sum_{L} k_{IL}~ e_L\bullet e_J
   \NN
   &=&\nabla^{\Sigma}_{e_I} e_J-\beta~\sum_{L,M} k_{IL}~ \epsilon_{LJM}~e_M
\ea
or in components using $g(e_I,e_J)=\delta_{IJ}$
\be
   A_{IJK}:=(A_I)_{JK}=g( \nablaA_{e_I} e_J,e_K)=\Gamma_{IJK}-\beta\sum_{L}k_{IL}~\epsilon_{LJK}~~.
\ee
Here $\Gamma_{KIJ}$ correspond to the usual Christoffel symbols\footnote{\label{Koszul}Its components can be obtained for an arbitrary basis from the well known Koszul formula \cite{Jost2008}:\linebreak
$
   \Gamma_{IJK}=g(\nabla^{\Sigma}_{e_I} e_J,e_K)
   =\frac{1}{2}\Big(~e_I\,g(e_J,e_K) - e_K\,g(e_I,e_J)+ e_J\,g(e_K,e_I)
   -g\big(e_I,[e_J,e_K] \big)+g\big(e_K,[e_I,e_J] \big)+g\big(e_J,[e_K,e_I] \big)~\Big)
$}.
\subsection{Curvature of the Ashtekar Connection}
Our previous considerations imply for the curvature of the Ashtekar connection 
\ba\label{curvature of nablaA}
   R^A(e_I,e_J)e_K
   &=& \nablaA_{e_I}\nablaA_{e_J} e_K 
                    - \nablaA_{e_J}\nablaA_{e_I} e_K 
                    - \nablaA_{[e_I,e_J]} e_K
   \NN  
   &=&\sum_M\Big\{\big(e_I(A_{JKM})-e_J(A_{IKM})\big)
            +\sum_L\big(A_{JKL}A_{ILM}-A_{IKL}A_{JLM}-C_{IJL}A_{LKM} \big)\Big\}~e_M               
   \NN               
   &=&R(e_I,e_J)e_K 
       +\beta \sum_{L,M}\Big\{\big(e_I(k_{JL})-e_J(k_{IL})\big)\epsilon_{KLM}
                               +\sum_N \big(\Gamma_{IJL}-\Gamma_{JIL}\big)k_{LN}\epsilon_{KNM}\Big\} ~e_M             
   \NN
   &&+\beta^2 \sum_M \big(k_{IK}\,k_{JM}-k_{IM}\,k_{JK}\big)~e_M                           
\ea
where $C_{IJL}=\Gamma_{IJL}-\Gamma_{JIL}=g([e_I,e_J],e_K)$ from the torsion freeness of $\nabla^{\Sigma}$.
This gives for the scalar curvature of $\nablaA$
\ba\label{scalar curvature of nablaA}
  R^A&:=&\sum_{I,J}g\big(R^A(e_I,e_J)e_J,e_I\big)
  \NN
  &=& \sum_{I,J}\Big\{\big(e_I(A_{JJI})-e_J(A_{IJI})\big)
            +\sum_L\big(A_{JJL}A_{ILI}-A_{IJL}A_{JLI}-C_{IJL}A_{LJI} \big)\Big\} 
  \NN
  &=& R+\beta^2 \sum_{IJ}\Big\{k_{IJ}\,k_{JI}- k_{II}\,k_{JJ}\Big\}
  \NN    
  &=& R-\beta^2 \Big\{ \big(\tr(k)\big)^2-\tr(k^2)\Big\}~~~,
\ea

where the term proportional to $\beta$ vanishes, because it is antisymmetric in $I,J$. This can be compared to the decomposition of the 4-dimensional curvature scalar under a $3+1$-decomposition $\M\cong\Sigma\times\mb{R}$ of the manifold \cite{Levermann2009}:
\ba\label{scalar curvature of nablaM}
   R^\M
   &=&R^\Sigma + s\Big(\big(tr(k)\big)^2-\tr(k^2)\Big)+2s\,\text{div}\big(-\tr(k)\,n+\nabla^\M_nn\big)
   \NN
   &=&R^A+(s+\beta^2)\Big\{\big(\tr(k)\big)^2-\tr(k^2)\Big\}+2s\,\text{div}\big(-\tr(k)\,n+\nabla^\M_nn\big)~~~.
\ea
Here $\text{div}(X)=\sum_I g(\nabla^\M_{e_I}X,e_I)+s~g(\nabla_n^\M X,n)$ denotes the divergence of a vector field $X\in \Gamma(T\M)$.
Starting with the Hamilton constraint given in \cite{Thiemann2007}, p 124 as
\ba
   H&=&-\frac{s}{\sqrt{\det q}}\Big(K^l_aK^j_b-K^j_aK^l_b \Big)E^a_j E^b_l-\sqrt{\det q}R
\ea
where $i,j,l=1,2,3$ denote $su(2)$-indices and $a,b$ denote indices in $T\Sigma$, moreover the densitized triads $E^a_i$ are given in terms of the non-densitized inverse triads $e^a_i$ as $E^a_i=\sqrt{\det q}~e^a_i$. This can be rewritten as
\be
   \frac{H}{\sqrt{\det q}}=-s\Big(\tr(K^2)-(tr(K))^2\Big) -R
\ee
Using (\ref{scalar curvature of nablaA}) this can be written as the well known form 
\be
   -\frac{\beta^2}{\sqrt{\det q}}H=s\,R^A+\big(\beta^2-s\big)R~.
\ee
This finishes the short digression on the coordinate free description of the Ashtekar variables. We will use this formalism in order to re-obtain the geometric meaning of the parameter $c$ in case of the homogeneous isotropic Bianchi I -universe.

\subsection{\label{Interpretation of the Symmetric Connection}Interpretation of the Symmetric Connection for Spatially Homogeneous Isotropic Models}
In what follows we will examine the properties of the Ashtekar-connection on Friedmann-Robertson-Walker-spacetimes, whose metric tensor can be written on $\M_I\cong I\times\Sigma$ ($I\subset\mb{R}$, open) as
\be
   g= s~dt\otimes dt+a(t)^2~g_{\Sigma}~~,
\ee 
where $(\Sigma,g_\Sigma)$ is a connected Riemannian manifold with constant sectional curvature $K_\Sigma=\kappa$ and $g_\Sigma$ is independent of $t$. At every $m\in\M_I$ we decompose $T_m\M_I$ into the tangent space $T_m\Sigma$ and an orthogonal direction spanned by the surface normal $ n=\frac{\partial}{\partial t}$. Assuming $[n,X]=0$ $\forall~X\in T_m\Sigma$, we have
\ba
   g\big(W_n(X),Y\big)
   &\stackrel{(\ref{matrix elements Weingarten map})}{=}&g(\nabla^\M_X n,Y)
   \NN
   &\stackrel{\text{footnote \ref{Koszul}}}{=}&\frac{1}{2}\frac{\partial}{\partial t}\Big(a(t)^2g_{\Sigma}(X,Y)\Big)
   \NN
   &=&a(t)~\dot a(t)~g_{\Sigma}(X,Y)
   ~~~~~~\leadsto~~W_n=\frac{\dot a(t)}{a(t)}~\text{Id}_{T\Sigma}~~.
\ea
Hence (\ref{action of nabla^A}) reads
\be
  \nabla^A_XY=\nabla^{\Sigma}_X Y + \beta~\frac{\dot a}{a}~ X\bullet Y
\ee
or in components as in (\ref{nablaA in ONB})
\ba
   A_{IJK}:=(A_I)_{JK}
   &=&g( \nablaA_{e_I} e_J,e_K)
   =\Gamma_{IJK}-\beta\sum_{L}k_{IL}~\epsilon_{LJK}
   \NN
   &=&\frac{1}{2} (C_{KJI}+C_{IJK}-C_{IKJ})-\beta~\frac{\dot a}{a}~\sum_L~\delta_{IL}\epsilon_{LJK}
\ea
where we have used the product structure $e_I\bullet e_J=\sum_{I,J}\epsilon_{IJK}\,e_K$.
Moreover  $C_{IJK}=g([e_I,e_J],e_K)$, and the result for $\Gamma_{IJK}$ follows directly from footnote \ref{Koszul}.
In case of Bianchi I we have $\Gamma_{IJK}=0$ and $g(\partial_m,\partial_n)=a^2\delta_{mn}$ hence the coordinate basis $\{\partial_m\}$ can be related to an orthonormal basis $\{e_I\}$ of $T\Sigma$  by $\partial_m=\sum_{m}e_{mI}~e_I=a \delta_{mI}~e_I $ .
Hence
$A_{IJK}=-\beta \frac{\dot{a}}{a}\sum_L\delta_{IL}\epsilon_{LJK}=(A_I^M\tau_M)_{JK}=-A_I^M\epsilon_{MJK}$. Therefore 
$A^M_I=\beta \frac{\dot{a}}{a} \delta^M_I$ and $A_b^M:=\sum_I e_{bI}A^M_I=\beta~a~\frac{\dot{a}}{a}\sum_I\delta_{bI}\delta^M_I=\beta~\dot{a}~\delta^M_b$.Therefore we get the usual \cite{Bojowald2008} identification
\be
   c=\beta\dot a
\ee

This yields for (\ref{curvature of nablaA}) in an orthonormal frame $\{e_I\}$ on $T\Sigma$
\be
   R^A(e_I,e_J)e_K=R^\Sigma(e_I,e_J)e_K + \beta^2 \left(\frac{ \dot a}{a}\right)^2\big(e_I\bullet e_J\big)\bullet e_K
\ee
and we consistently obtain for Bianchi I with $R^\Sigma=0$ (\ref{scalar curvature of nablaA})
\be
   R^A=-6\beta^2 \left(\frac{ \dot a}{a}\right)^2
      ~~~~\Leftrightarrow~~~~
   {c^2}=-\frac{a^2}{6}~ R^A
        \stackrel{(\ref{scalar curvature of nablaA})}{=}
        \frac{a^2}{6}\beta^2 \Big\{ \big(\tr(k)\big)^2-\tr(k^2)\Big\}
        =(\beta\dot{a})^2
\ee
where we have used the fact that $k_{IL}=-\frac{\dot a}{a}\delta_{IL}$ and $\dim\Sigma=3$.
Evaluating (\ref{scalar curvature of nablaM}) for Bianchi I we find in agreement to \cite{Wald1984}, p 97:
\be
  R^\M=-6\,s\,\left(\Big(\frac{\dot a}{a}\Big)^2+\frac{\ddot a}{a}\right)~.
\ee
This shows, that in a situation where $\ddot a\ll \dot a$ one can understand the limit $c\rightarrow -\infty$ as a blow up of scalar 4-curvature on $\M$, that is a situation close to a singularity. The fact that in the limit $c\rightarrow\infty$ we re-obtain the LQC framework indicates an affirmation of the BKL-picture, as discussed in section \ref{Asymptotic Analysis and the BKL Picture} .

\end{appendix}
\pagebreak
\addcontentsline{toc}{section}{\numberline{}References}

\end{document}